\documentclass[fleqn,usenatbib]{mnras}

\usepackage{newtxtext,newtxmath}

\usepackage[T1]{fontenc}

\DeclareRobustCommand{\VAN}[3]{#2}
\let\VANthebibliography\thebibliography
\def\thebibliography{\DeclareRobustCommand{\VAN}[3]{##3}\VANthebibliography}

\usepackage{graphicx}	
\usepackage{amsmath}	
\usepackage{comment}

\newcommand{\msun}{\mathrm{M}_\odot}

\newcommand{\subfind}{\textsc{Subfind}}
\newcommand{\sextractor}{\textsc{SExtractor }}
\newcommand{\modtext}{\color{black}}
\newcommand{\modtextt}{\color{black}}

\title[Diffuse light measurements in galaxy groups]{How to Interpret Measurements of Diffuse Light in Stacked Observations of Groups and Clusters of Galaxies}

\author[S. L. Ahad et al.]{Syeda Lammim Ahad,$^{1}$\thanks{E-mail: ahad@strw.leidenuniv.nl}
Yannick M. Bah\'{e},$^{1}$
Henk Hoekstra$^{1}$
\\
$^{1}$ Leiden Observatory, Leiden University, P.O. Box 9513, 2300 RA Leiden, The Netherlands\\
}

\date{Accepted XXX. Received YYY; in original form ZZZ}

\pubyear{2022}

\begin{document}
\label{firstpage}
\pagerange{\pageref{firstpage}--\pageref{lastpage}}
\maketitle

\begin{abstract}
The diffuse light within galaxy groups and clusters provides valuable insight into the growth of massive cosmic structures. Groups are particularly interesting in this context, because they represent the link between galactic haloes and massive clusters. However, low surface brightness makes their diffuse light extremely challenging to detect individually. Stacking many groups is a promising alternative, but its physical interpretation is complicated by possible systematic variations of diffuse light profiles with other group properties. Another issue is the often ambiguous choice of group centre. We explore these challenges using mock observations for 497 galaxy groups and clusters with halo masses from $\sim 10^{12} \textrm{M}_{\odot}$ to $1.5 \times 10^{15}\textrm{M}_{\odot}$ at redshift $0.1$ from the Hydrangea cosmological hydrodynamic simulations. In 18 per cent of groups with at least five galaxies above $10^{9} \msun$ in stellar mass, the $r$-band brightest galaxy is not the one at the centre of the gravitational potential; line-of-sight projections account for half of these cases. Miscentring does not significantly affect the ensemble average mass density profile or the surface brightness profile for our sample: even within ambiguously centred haloes, different centring choices lead to only a 1 per cent change in the total fraction of diffuse intra-group light, $f_{\textrm{IGL}}$. We find strong correlations of $f_{\textrm{IGL}}$ with the luminosity of the central group galaxy and halo mass. Stacking groups in narrow bins of central galaxy luminosity will therefore make the physical interpretation of the signal more straightforward than combining systems across a wide range of mass.
\end{abstract}

\begin{keywords}
galaxies: groups: general -- galaxies: clusters: general -- galaxies: evolution -- galaxies: stellar content -- methods: numerical
\end{keywords}



\section{Introduction}

According to the hierarchical structure formation model from the $\Lambda$-cold dark matter ($\Lambda$CDM) cosmological paradigm, large-scale structures in the Universe, such as galaxy groups and clusters, assembled via the merging and accretion of smaller systems. During this assembly, the tidal stripping of stars from galaxies and the accretion of smaller systems by the central galaxy produces a diffuse stellar component that surrounds the dominant galaxies in groups and clusters of galaxies. This is most clearly visible around the brightest galaxies in massive clusters (BCGs), where the diffuse low surface brightness light is more commonly referred to as the intracluster light (ICL). The radial extent of ICL around BCGs and the contribution of the ICL in the total light from groups and clusters can provide important constraints for cosmic structure formation. As this diffuse light can extend out to hundreds of kilo-parsecs from the cluster centre and often envelops multiple galaxies in the cluster, it is commonly considered a separate component of the galaxy groups and clusters \citep[for recent reviews, see, e.g.][]{Mihos2015,Contini2021,Montes2022}. 

In recent years, there have been increasing efforts to study the ICL in clusters, both by using high-quality data for individual clusters \citep[e.g.][]{Mihos2005, Montes2014,Montes2018, JimenezTeja2018,Demaio2018,Demaio2020,Montes2021}, or by stacking a statistical sample of groups/clusters to improve the signal-to-noise-ratio \citep[SNR, e.g.][]{zibetti2005,Zhang2019}. Depending on the methods used to separate the ICL from the BCG light, the ICL can comprise more than 30\% of the total star light of the host cluster \citep[e.g.][]{zibetti2005,Gonzalez2013,Mihos2017,Montes2018,Zhang2019,Kluge2021}, although a consensus about the ICL fraction is yet to be reached from both the simulation and observation sides \citep[see, e.g. table 1 from][]{Kluge2021}. Through a large number of simulation studies, the origin and growth of the ICL have been attributed to multiple channels (see e.g.~\citealt{Mihos2017} and \citealt{Contini2021} for more discussion on the origin and growth of the ICL)
, including galaxy mergers \citep{Murante2007}, tidal stripping \citep{Gallaghar1972}, galaxy disruption \citep{Guo2011}, and even in-situ star formation in the intracluster medium \citep{Puchwein2010,Tonnesen2012}. Some recurring findings from recent studies are that the ICL mass distribution follows the global dark matter (DM) distribution, both in observations \citep[e.g.][]{Montes2019} and simulations \citep[e.g.][]{AlonsoAsensio2020}, and that ICL profiles align more with the underlying cluster halo than with the BCG \citep{Kluge2021}, making the ICL an indirect probe for tracing the build-up of the structures. These findings collectively confirm that the ICL growth is indeed connected to the evolution of the large elliptical galaxies such as the BCG, the baryon fraction of galaxy clusters, and the build-up of large scale structures like galaxy clusters where the giant galaxies (BCG) reside.  

Although most works are based on clusters, because the ICL is more prominent within these most massive haloes and clusters are preferentially targeted by deep surveys such as the Hubble Frontier Fields (HFF) \citep{Lotz2017} or BUFFALO \citep{steinhardt2020}, studying the diffuse light in groups, or `intragroup light' (IGL) is particularly interesting for several reasons. Groups cover the intermediate halo mass regime of cosmic structures between galaxy-mass haloes and galaxy clusters. They are also the main building blocks of clusters. Therefore, understanding the build-up of the diffuse stellar component in groups will improve our understanding of the growth of ICL in clusters. Also, compared to clusters, groups are dynamically less disturbed, have had fewer interactions with other systems, and are more concentrated. As a result, it is more straightforward to connect the growth of the IGL in groups with their dynamical history. 


Even though the importance of understanding the buildup of IGL/ICL across a wide range of host halo mass has been recognized for a while, there have only been a few studies on the diffuse light in a large enough sample of group-mass haloes so far \citep{zibetti2005,poliakov2021}. From the observational side, the main reason behind this is the lack of a reliable group catalogue with large enough sample size. \citet{zibetti2005} studied the diffuse light in 683 SDSS groups and clusters by stacking them to increase the SNR and found that the surface brightness of the diffuse light correlates with BCG luminosity and with cluster richness, but the fraction of the total light in the diffuse component does not vary notably with these properties. However, they only studied these behaviours by dividing their sample in two sub-samples for each property which may not be representative of the full variation of these properties. The dependence of the diffuse light fraction on different group properties (e.g. group halo mass, richness) therefore still remains an open question and needs to be studied with multiple approaches to be well-understood. With high-quality multi-band ($u, g, r, i$) photometry of the Kilo Degree Survey \citep[KIDS,][]{dejong2013} and a group catalogue \citep{Robotham2011} based on the highly complete spectroscopic Galaxy and Mass Assembly \citep[GAMA,][]{Driver2009,driver2011} survey, it is now possible to push the detection limit of the diffuse light into group-mass haloes and improve the interpretation of the data. 

Studying the light distribution of individual groups is useful to understand the diversity of the IGL signal and their formation channels. However, the low surface brightness of the IGL means that individual systems have a very low SNR, which results in a higher uncertainty in the interpretations. Stacking the light of multiple groups can help to improve the SNR while keeping the key features of the underlying population. 
However, before simply stacking all the group data, we need to consider a few caveats. In previous group catalogues, the group centres were determined by either taking the centre of the distribution of light, or the brightest galaxy, or the galaxy with the highest stellar mass in the system. Stellar-to-halo-mass-relations (SHMR) have also been used to utilize the halo mass to determine the physical halo centre, but they were mainly assuming a monotonic relation that assigned a fixed halo mass for a fixed galaxy luminosity irrespective of their colour (see e.g. sec. 1.1 of \citealt{tinker2020} for more discussion on this). However, in recent years, it has been shown that the SHMR for central galaxies depends on their colour: a bluer central typically resides in a less massive halo compared to a red central of the same stellar mass \citep[e.g.][]{Bilicki2021,Mandelbaum2016}. This calls for a re-estimation of the group centres in existing group catalogues, especially the ones that were based on only stellar mass or luminosity. The obvious question is whether this improves our estimates of the group centres. Specifically, how can we determine a spurious central estimation and adjust for any biases that are introduced by such miscentring? And if we are indeed misidentifying an appreciable fraction of group centres, how much does it affect the IGL estimation? Exploring these questions is particularly important to ensure that stacking will actually improve the SNR -- not make it worse by adding unwanted signals from miscentred groups and that analysing the stacked profile will provide us with unbiased interpretations regarding the IGL.

Another issue is the dependence of variation in the IGL/ICL distribution on the properties of the central galaxy and the host system. \citet{ContiniGu2021} reported that the ICL fraction and distance from the centre to where ICL dominates the total galaxy mass vary widely (from 15~kpc to 100~kpc) depending on the morphology (bulge or disk dominated) and dynamical history of the BCG. \citet{Kluge2021} found a positive correlation between BCG+ICL brightness and the host cluster mass, cluster size (radius), and integrated light in the satellites. Therefore, while stacking the group profiles, it is necessary to quantify the effect of galaxy properties and to find the optimal way of scaling and stacking to ensure a robust interpretation of the profiles. 

The arguably easiest way to explore this is by using cosmological hydrodynamic simulations. In recent years, their fidelity has improved enough to help us understand such intricate details as the IGL \citep[see e.g.][for a recent review]{oppenheimer2021}. The Hydrangea simulation suite of 24 massive galaxy clusters \citep[][]{bahe2017hydrangea} is an excellent sample to study the IGL around groups because they are specifically made to study galaxy evolution in and around large scale structures. The simulations have also been successful in reproducing the stellar mass distribution of satellite galaxies in both the local Universe \citep{bahe2017hydrangea,barnes2017cluster} and intermediate to high redshift \citep[$0.6<z<2.0, $][]{ahad2021} for galaxy clusters. 

In this paper, we explore the impact of miscentring in galaxy groups on the IGL measurements in them {\modtext with the analysis of simulated data (Hydrangea) that are matched to an observational dataset (GAMA).} We also study the extent of the IGL around the group centre to determine the best way of stacking galaxy groups to interpret the IGL signal. 

The organization of the paper is as follows. In Sec.~\ref{sec:gama_kids_data}, we describe the Galaxy And Mass Assembly (GAMA) survey data, our group sample selection from the KiDS+GAMA overlap, and test how using a colour-dependent SHMR affects the group centre selection for the GAMA group sample. We describe the Hydrangea simulation suite in Sec.~\ref{simdata}. Here, we also discuss the group sample selection from the simulations and preparation of the mock photometric data. In Sec.~\ref{sec:miscentring_analysis}, we explore the effects of possible miscentring on the IGL measurements by means of the density profiles and the surface brightness profiles of the simulated sample. We discuss the dependence of the measured IGL fraction on different properties of the brightest group galaxy (BGG) and host group properties, and the radial $u-r$ colour profile of the BGG+IGL in Sec.~\ref{sec:icl_on_bcg_morphology}. Finally, we summarize our conclusions in Sec.~\ref{sec:conclusions}.

\section{GAMA group data}

\label{sec:gama_kids_data}
\subsection{Galaxy and Mass Assembly survey data}
\label{sec:gama_data_intro}

The Galaxy And Mass Assembly (GAMA) galaxy survey \citep{Driver2009,driver2011,driver2022} is a unique project with 21-band photometric data and spectroscopic redshifts of $\sim 300,000$ galaxies. It is 98.5\% complete for SDSS-selected galaxies with $r < 19.8$ mag.
The spectroscopic survey was conducted using the AAOmega multi-object spectrograph on the Anglo-Australian Telescope to measure galaxy spectra in five fields covering a total of $\sim$ 286 deg$^2$ area, which provided detailed redshift sampling. 

In our work, we used the latest GAMA-II Galaxy Group Catalogue \citep[G$^3$CFOFv08,][]{Robotham2011}, generated using a friends-of-friends (FOF) based grouping algorithm in which galaxies are grouped based on their line-of-sight and projected physical separations, and the accompanying galaxy catalogue G$^3$CGalv09 \citep{Liske2015}. The catalogue consists of 23654 groups across all the GAMA fields and $N_{\textrm{FOF}}\geq2$ spectroscopically confirmed member galaxies. To ensure the most robust group selection, we only considered groups with $N_{\textrm{FOF}}\geq5$ here. We also used the stellar mass estimates as well as $u-$ and $r-$band magnitudes of GAMA galaxies 
from the StellarMassesLambdarv20 catalogue \citep{Taylor2011,Wright2016}, which includes physical parameters based on stellar population fits to rest-frame $ugrizY$ SEDs, and matched aperture photometry measurements of SDSS and VIKING photometry for all the $z < 0.65$ galaxies from the GAMA Panchromatic Data Release \citep{Driver2016}. This sample contains over 198,000 galaxies, with a median $\log(M_{\star}/$M$_{\odot}) \approx 10.5$ assuming $H_0 = \textrm{70km s}^{-1} \textrm{Mpc}^{-1}$. Further details on the GAMA stellar mass derivation can be found in \citet{Taylor2011} and \citet{Wright2016}.

Four of the GAMA fields (equatorial G09, G12, and G15 of 60 deg$^2$ each, and southern G23 of 51 deg$^2$) entirely overlap with the Kilo Degree Survey \citep[KiDS,][]{dejong2013} -- a large, deep, multi-band optical imaging survey that is designed for measuring cosmological parameters and covers 1350 deg$^2$ in four broadband filters $(u,g,r,i)$. The GAMA group catalogue, accompanied by the deep KiDS imaging (mean limiting $m_r = 25.02$ at $5\sigma$ significance in a $2''$ aperture, mean FWHM of the $r-$band PSF 0$\farcs{7}$, sampling 0$\farcs{213}$/pixel) from data release 4 \citep[DR4,][]{Kuijken2019} provides us with a unique opportunity to analyse the IGL around the low-mass galaxy groups, which we present in a companion paper (Ahad et al., in prep.). In order to make our results from this work more applicable to the IGL measurement in GAMA groups with KiDS imaging, we considered only GAMA+KiDS cross-matched groups in our sample. With an additional $N_{\textrm{FOF}}\geq5$ selection cut, we obtained a final sample of 2385 groups which we used in this work.

\subsection{Challenges in identifying the central galaxy in groups}
\label{subsec:bcg_kids_gama}

In the GAMA group catalogue (G$^3$CFOFv08), {\modtext plausible central galaxy candidates in each group} are selected in three ways \citep{Robotham2011}: (i) taking the galaxy at the centre of light (CoL) distribution; (ii) taking the brightest galaxy in the group; and (iii) with an iterative method that starts by taking the group light distribution and then successively discards the galaxy that is the farthest from the CoL. The process is iterated until only two galaxies are left in the group, after which the brighter one is chosen as the central galaxy. After comparing to a mock galaxy catalogue obtained from populating the Millenium dark-matter simulations \citep{springel2005cosmological} with galaxies using the \textsc{GALFORM} \citep{Bower2006} semi-analytic model, \citet{Robotham2011} concluded that the iterative method provided the most robust selection of the centrals and recommended the selected galaxy with this method as the {\modtext optimal group centre candidate}. Throughout this paper, these iterative centres from the GAMA group catalogue are referred to as the G3C centrals.

Selecting the centre of light or the brightest galaxy in the halo provides a plausible estimate of the halo centre in most cases. This is because the brightest galaxy is typically expected to have more stars (as a result, more mass) than the rest of the group galaxies, and therefore is located at the centre of potential of the halo. But this is not always the case -- the mass-to-light ratio is different based on galaxy colours. This means that at a fixed luminosity, a bluer galaxy will have lower stellar mass compared to a redder galaxy \citep[e.g.][]{vandesande2015,garciabenito2019}. Moreover, at a fixed stellar mass, the stellar-to-halo-mass ratio is also observed to be different for red and blue galaxies. Using weak lensing measurements for the halo mass of the bright galaxy sample from the KiDS (which include the central galaxies from the KiDS+GAMA overlap), \citet{Bilicki2021} studied the SHMR separately for blue and red galaxies. 
They reported that for the same stellar mass ($M_{\star}$), redder galaxies typically reside in a more massive halo compared to the bluer ones (see their fig.9). The difference can be a factor of two at $M_{\star}<5\times10^{10}h^{-2}$M$_{\odot}$, which is the mass of most of the group centrals in our sample, and even more at higher $M_{\star}$s. Combined with the colour-dependent mass-to-light ratio, this implies that even if a blue central candidate emits more light, it may have a lower total (baryon + DM) mass than a slightly less massive red central candidate and therefore may not actually sit at the centre of potential of the group halo. This insight calls for a revisit of the GAMA group catalogue to see how robust the G3C central selection is when accounting for a colour-dependent SHMR.

\begin{figure}
	\includegraphics[width=\columnwidth]{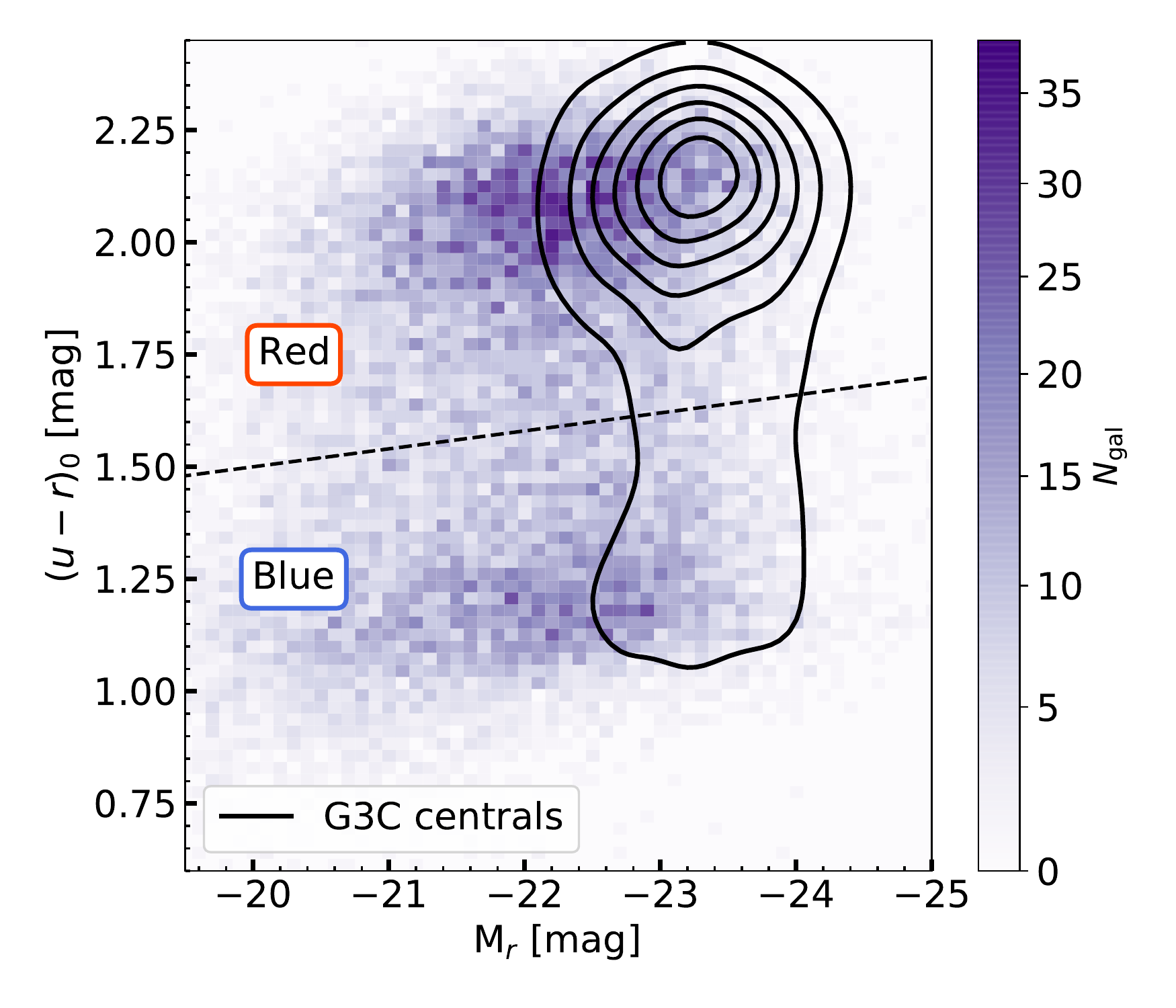}
	\caption{The rest frame $(u-r)$ colour of all the GAMA galaxies in our sample (purple). Black contours show the Gaussian kernel density estimation of the rest frame $(u-r)$ colour of the GAMA group centrals in our sample. The black dashed line is used to separate the sample into red (above) and blue (below) galaxies. As expected, the centrals are indeed the brightest galaxies in the entire sample. A non-negligible fraction (28\%) of them are blue, which are conceivably misidentified as centrals (see text).}
	\label{fig:ur_vs_mr_gama}
\end{figure}

Following the example of fig. 6 of \citet{Bilicki2021}, we examined the distribution of GAMA group galaxies (both satellites and centrals) in the rest-frame $(u-r)$ colour vs $r$-band magnitude parameter space (all magnitudes are obtained from the StellarMassesLambdarv20 catalogue), and separated them into blue and red samples. Figure~\ref{fig:ur_vs_mr_gama} shows the 2D histogram of the GAMA group galaxies in purple. A clear separation in colour is evident from the distribution, divided by a straight line with slope -0.04 and intercept 0.7 (black dashed line) to separate the red galaxies (above) from the blue (bottom) ones. This separating line qualitatively agrees with that from \citet{Bilicki2021}, although they used a different colour ($u-g$) to separate their sample into red and blue galaxies. The distribution of the G3C centrals is shown by the black contours, which are the normalized Gaussian kernel density estimates of the centrals. Unsurprisingly, these are located at the brightest end of the galaxy distribution. More noteworthy, however, is the fact that, although most centrals are ``red'' according to our definition, a non-negligible minority (28\%) lie below our dividing line and are hence classified as ``blue''. 

Using the colour distribution of the GAMA galaxies, we selected the three brightest galaxies ($r$-band magnitude) from each of the GAMA groups in our sample and assigned a halo mass to each of them by inverting the best-fit SHMR for the appropriate colour found by \citet{Bilicki2021}(their eqn. 7):
\begin{equation}
    M_{\star} (M_{\rm{h}}) = M_0 \frac{(M_{\rm{h}}/M_1)^{\gamma_1}}{[1 + (M_{\rm{h}}/M_1)]^{\gamma_1 - \gamma_2}}
\end{equation}
Here, $M_{\star}$ is the stellar mass, $M_{\rm{h}}$ is the corresponding halo mass, and $M_0$, $M_1$, $\gamma_1$, and $\gamma_2$ are {\modtext constants that are assigned different values depending on whether the galaxy in consideration is blue or red} (see table 3 of \citealt{Bilicki2021}).

\begin{figure}
	\includegraphics[width=\columnwidth]{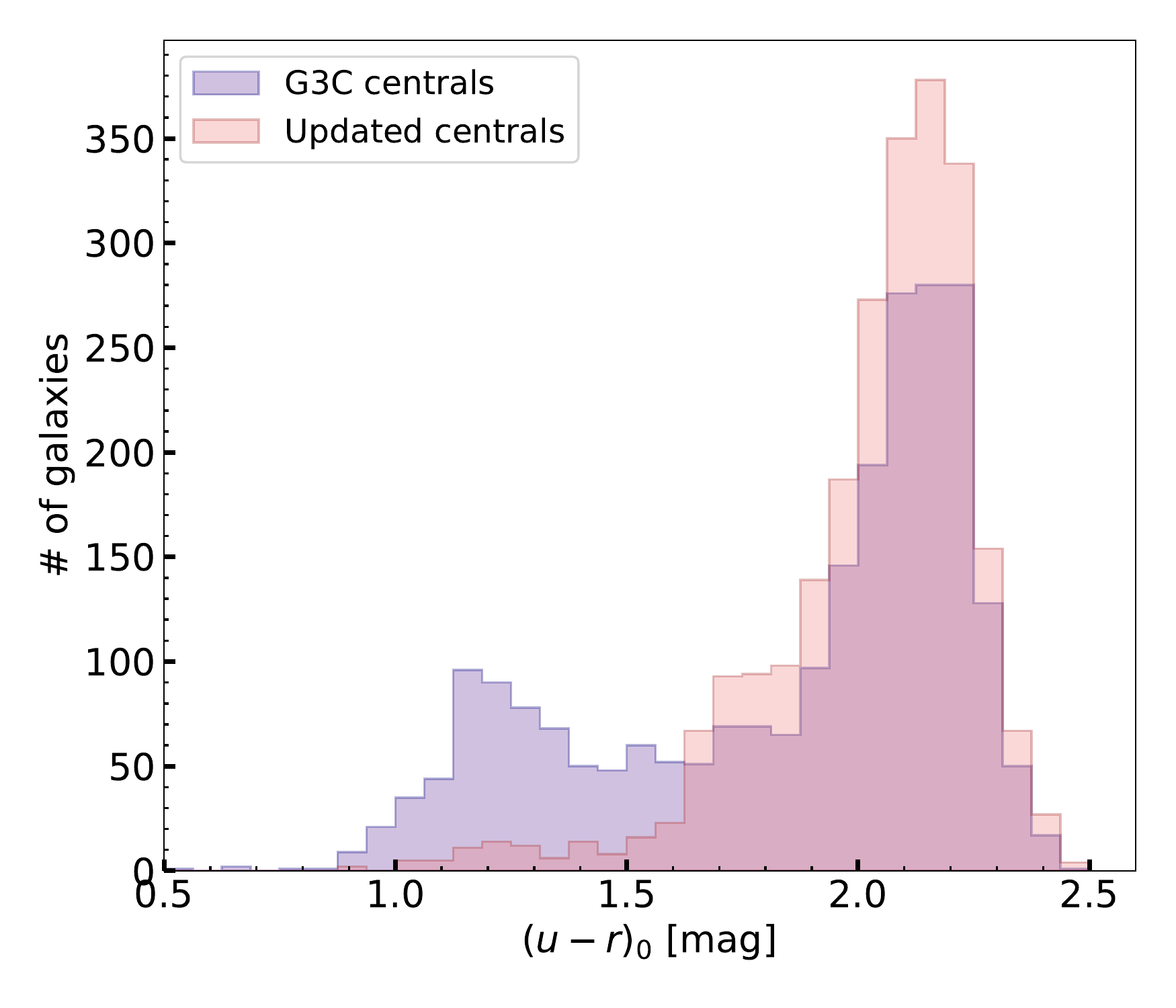}
	\caption{Distribution of the rest-frame $(u-r)$ colour of central galaxies from the original GAMA catalogue (purple) and our updated centrals using a colour-dependent SHMR (orange). Compared to the GAMA catalogue, the fraction of blue BGG candidates ($u-r\lesssim 1.6$) is strongly reduced in our updated sample. These groups have been re-assigned with a red central with a higher halo mass.}
	\label{fig:Gama_updated_colour_hist}
\end{figure}

Only the brightest three galaxies from each group were considered to remain within the brightness and stellar mass range considered by \citet{Bilicki2021}. {\modtext We picked the galaxy with the highest predicted halo mass as the new central galaxy candidate for the groups because
the galaxy with the highest halo mass is likely to have the deepest gravitational potential, and consequently to reside at the centre of potential of the host group.} The rest-frame $(u-r)$ colour distribution of the central galaxy candidates from the original GAMA catalogue and our updated sample are shown in Fig.~\ref{fig:Gama_updated_colour_hist} in purple and orange histograms, respectively. Out of the 2385 groups in our sample, the fraction of blue centrals has decreased from 28\% to 4.4\% in our updated central galaxy candidates. Repeating the same procedure but this time selecting the three galaxies with the highest stellar mass from each GAMA group in our sample, resulted in a similar central galaxy re-assignment. However, the validity of these re-assigned central candidates must be checked before drawing any definite conclusion on whether it actually provides a physically more robust group centre in ambiguous cases. We do so in Sec. \ref{sec:miscentring_analysis}.

Another uncertainty in the BGG assignment may come from the aggregation (fragmentation) of low-mass (high-mass) groups in the FOF halo finder algorithm: \citet{Jakobs2018} reported that aggregation of multiple low mass groups into one may cause the halo finder to introduce BGG candidates that are not physical members of the group, and fragmentation of high-mass groups may cause the halo finder to pick up the right BGG but for a halo mass that is too low. They found such effects present in 37\% groups/clusters in their sample. Although the fraction of groups where this fragmentation/aggregation effect may occur is unknown for our sample, different small effects can result in a non-negligible accumulated impact during the analysis. We therefore next investigate the miscentering and its impact on diffuse light on simulated galaxy groups.



\section{Simulated data}
\label{simdata}
\subsection{The Hydrangea Simulation Suite}
\label{hydrangea}
The Hydrangea simulations \citep[part of the C-EAGLE project,][]{bahe2017hydrangea, barnes2017cluster} are a suite of high-resolution cosmological hydrodynamic zoom-in simulations of $24$ massive galaxy clusters. The simulation regions were chosen from a low-resolution dark matter(DM) only parent simulation \citep{barnes2017redshift} of a (3200~co-moving Mpc$)^3$ volume. Each of the high-resolution simulation regions is centred on a massive cluster ($10^{14.0} \leq M_{200c}/\msun \leq 10^{15.4}$ at $z = 0$)\footnote{$M_{200c}$ refers to the mass enclosed within a sphere centred at the potential minimum of the cluster radius $r_{200c}$, within which the average density of matter equals 200 times the critical density.}. The particle mass resolutions are $m_\textrm{baryon} =1.81 \times 10^6\,\msun$ and $m_\textrm{DM} = 9.7 \times 10^6\,\msun$; the gravitational softening length is $0.7$ physical kpc (pkpc) at $z<2.8$. The high-resolution simulation regions include the large scale surroundings of the clusters to $\geq 10$ virial radii ($r_{200c}$) at $z=0$, and therefore contain many group scale haloes, in addition to the central clusters. The simulations were run using the AGNdT9 calibration of the EAGLE galaxy formation and evolution code (for details about the simulation model, hydrodynamics scheme, and comparison of the model to observed galaxy properties, see \citet{schaye2014eagle, schaller2015eagle, crain2015eagle}  and references therein). The subgrid physics models used to simulate astrophysical processes that originate below the resolution scale of the simulation include star formation (following \citealt{schaye2008relation}, with metallicity-dependent star formation threshold from \citealt{schaye2004star}), star formation feedback \citep{dalla2012simulating}, radiative cooling and heating \citep{wiersma2009effect}, stellar evolution \citep{wiersma_2009_metallicity}, black hole seeding, growth, and feedback (\citealp{rosas2015impact,schaye2014eagle}; see also \citealp{bahe2022})
A flat $\Lambda$CDM cosmology is assumed in the simulations  with parameter values $H_0$ = 67.77 kms$^{-1}$Mpc$^{-1}$, $\Omega_{\Lambda} = 0.693$, $\Omega_\textrm{M} = 0.307$, and $\Omega_\textrm{b} = 0.04825$ \citep{planck2013}. We also use the same cosmology throughout any relevant calculations in this paper. Although this slightly differs from the cosmological parameters used for the GAMA catalogue calculations (mentioned in Sec. \ref{sec:gama_data_intro}), our conclusions are not affected. 

The primary output of each simulation consists of 30 snapshots between $0<z<14$ with a time step of
500 Myr. In each of these snapshots, gravitationally bound structures (and stellar, DM, and gas content of each object) were identified with the \subfind{} code \citep{springel2001,dolag2009}, through a two-step process. In the first step, a friends-of-friends (FOF) algorithm was used to identify spatially disconnected groups of DM particles.
Baryon particles were connected to the FOF group (if any) of their nearest DM particle \citep{dolag2009} and only FOF groups with more than 32 DM particles were considered. The following step selected gravitationally bound candidate `subhaloes' within each FOF group as locally over-dense regions. Particles in the FOF group that are not bound to any subhalo, but still gravitationally bound to the group and are self-bound to each other, were considered as the `background' or `central' subhalo (see \citealt{bahe2017hydrangea} and \citealt{Bahe_et_al_2019} for more details). In this paper, we refer to all the subhaloes other than the central subhaloes as the `satellites'.  
An important distinction is that in the simulations the central subhalo in the FOF group contains all self-bound particles that are not in a satellite subhalo. Hence, it comprises what would observationally be described as the combination of the central galaxy and the IGL, without any explicit distinction between the two.
In Sec.~\ref{ss:bcg_icl_sep}, we discuss different approaches to separate the BGG from the IGL that are directly comparable to observational methods and whether the difference in methods can be quantitatively connected to the detected IGL fraction.

In this work, we have used both the particle data and the FOF groups and subhaloes from \subfind{} outputs at $z = 0.101$. The redshift was chosen to match our group selection from the GAMA group sample described in Sec. \ref{sec:gama_kids_data}. For the analysis, we prepared mock $r-$band images centred on each of the FOF groups in our sample using projected particle luminosities \citep{Negri2022} that have been k-corrected to $z=0$ following \citet{chilingarian2010} and \citet{chilingarian2012}. The size of each of the images is 2 pMpc along both axes. The pixel-to-arcsecond ratio of 0.213 and an appropriate {\modtext RMS noise (pixel value $\sim 10^{-12}$, in units
of flux relative to the flux corresponding to magnitude = 0, this is the same noise level given the zero-point of the AstroWise pipeline that was used to process the KiDS DR4 data}) 
was applied to mimic the KiDS images from DR4 \citep{Kuijken2019}. One major advantage of using simulated data compared to the observational data is that we can also create mock observations of only the central subhalo within each FOF group -- recall that this includes the IGL -- and analyse its light without the need to mask satellites or line-of-sight projections. We utilized this and prepared another set of mock $r-$band images using only the particles of the central subhalo. All the other specifications of these images remained the same as the projected group images described above. 

There are a few points of concern while using the simulated data for our analysis. Biases can be introduced due to the failure of \subfind{} to assign star particles to satellites (Bah\'{e} et al., in prep). Uncertainties in the produced IGL fraction and its radial distribution may arise from the star formation rate model and the resolution limit of the simulations. {\modtext The EAGLE model matches the observed stellar mass and luminosity functions and their evolution considerably well up to $z=7$ \citep{schaye2014eagle,furlong2015,Trayford2015}. The Hydrangea simulations can also reproduce the observed stellar mass functions and mass density profiles in galaxy clusters out to $z=2$ \citep{bahe2017hydrangea,ahad2021}. However, an offset in the observed size-mass relation towards more compact passive galaxies in the AGNdT9 variant of the EAGLE model \citep{schaye2014eagle} may have been responsible for a lower mass fraction of ICL in the central clusters from the Hydrangea simulations compared to the observed ICL mass fractions as compact galaxies are less prone to stripping \citep{AlonsoAsensio2020}.} We note that the opposite effect is described in \citet{Henden2020}, who found that boosted tidal stripping from artificially large satellite galaxies increases the ICL stellar mass fraction in their simulations. They also noted that uncertainties in galaxy sizes are the principal contributors to the uncertainty in determining the ICL mass fraction in simulations. In this work, we focus on direct observable properties based on mock observations of multi-band photometry. However, discrepancies in the IGL mass fraction measurements caused by the above-mentioned attributes of the simulations may add uncertainties to the optical measurements of the IGL fractions.

\subsection{Group Selection}
\label{ss:hydrangea_group_sel}
We performed a detailed selection procedure to match the simulated groups with the GAMA group sample (described in Sec.~\ref{sec:gama_kids_data}). 
We considered only the simulated data at $z=0.101$ and compared this to a matched subsample from our GAMA group sample at $0.09<z<0.15$ to minimise any potential redshift evolution within the observations. A detailed analysis of the redshift dependence of the IGL will be discussed in a follow-up paper. We excluded groups near the edge of the simulation zoom-in region to avoid numerical artefacts caused by the artificial gas vacuum and resolution jump outside of the high-resolution region. To match our GAMA sample selection, we only kept groups with at least five member galaxies {\modtext brighter than an $r-$band magnitude limit of 19.8}. For this, we took the absolute magnitudes of the galaxies in $r-$band within 30~kpc aperture, applied the appropriate k-correction to the magnitudes at $z=0.101$, and applied the distance modulus to obtain the final apparent magnitudes of the group galaxies. {\modtext The $r-$band absolute magnitude of the central galaxies were also computed from the particle magnitudes within 30~kpc aperture.} In our final sample, we have a total of 497 groups, including the 24 central clusters. We prepared a catalogue that contains details (e.g. halo mass, virial radius, BGG ID and location, colour and magnitudes, stellar mass) for these 497 groups to use in our analysis and refer to this as the `Hydrangea group catalogue' from now on.

\begin{center}
    \begin{figure*}
    	\includegraphics[width=1\textwidth]{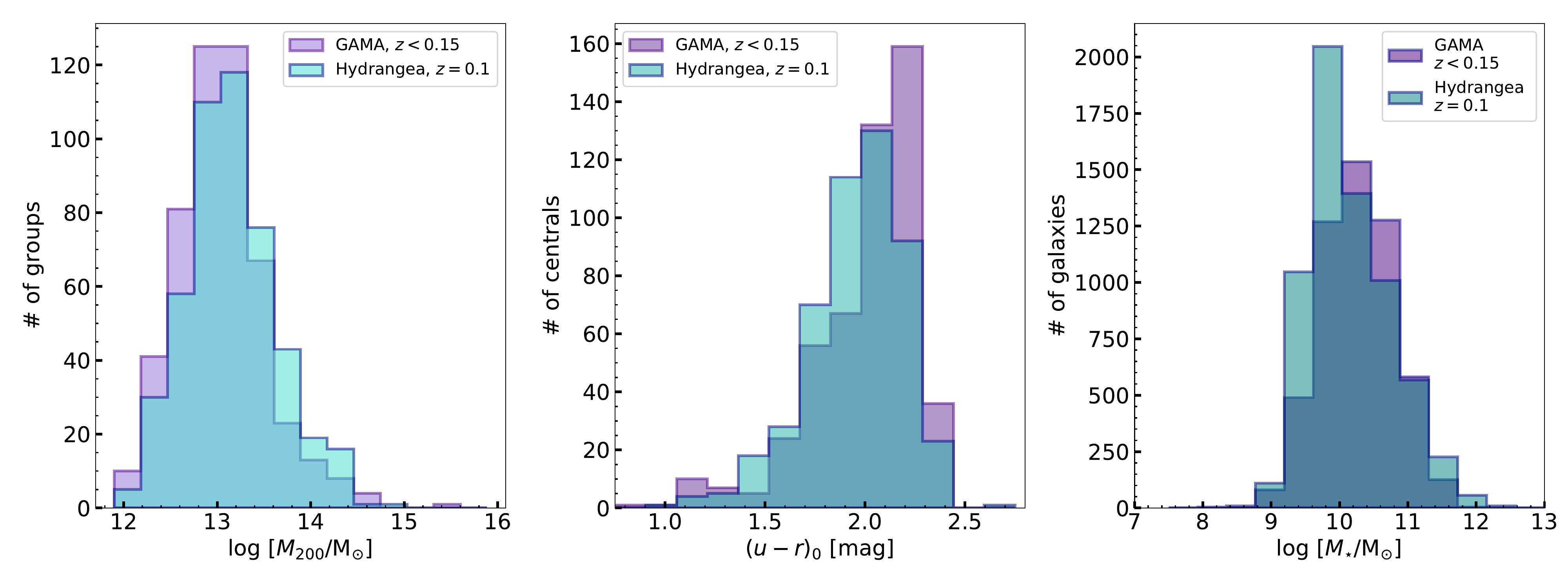}
    	\caption{Histograms comparing the distribution of different properties of the GAMA (purple) and Hydrangea (turquoise) group samples at comparable redshift ($0.08<z<0.15$ for GAMA; $z=0.1$ for Hydrangea). The left, middle, and right panels show the distributions of halo mass $(\log[{M}_{200} /\textrm{M}_{\odot}])$, dust-corrected rest-frame $(u-r)$ colour of the group BGGs, and total stellar mass of all the group galaxies in the GAMA and Hydrangea samples, respectively. The observed and simulated samples have comparable distributions in these three fundamental properties.
    	}
    	\label{fig:gama_hydrangea_samp_comp}
    \end{figure*}
\end{center}

Figure~\ref{fig:gama_hydrangea_samp_comp} shows the comparison between the GAMA group sample and the final Hydrangea group sample (excluding the central clusters in each simulation volume). The left, middle, and right panels show the group halo mass $(\log [{M}_{200} /\textrm{M}_{\odot}])$ distribution, the distribution of the rest frame $(u-r)$ colour of the group BGGs, and the stellar mass distribution of all the group galaxies {\modtext (both centrals and satellites)} from the GAMA (purple) and Hydrangea (turquoise) samples, respectively. The $(u-r)$ colour is obtained from dust-corrected magnitudes of \citet{Negri2022}. 
The samples show an excellent agreement in all the panels, {\modtext especially given that we do not expect perfect agreement between the samples because the Hydrangea sample is not mass-complete.} The match suggests that the simulated data can be used to test and predict different properties of the observed group data sample. 

\section{Effect of miscentring on the IGL fraction}
\label{sec:miscentring_analysis}

We found in Sec. \ref{subsec:bcg_kids_gama} that 23\% of the GAMA groups have a reassigned (redder) central galaxy based on the colour-dependent SHMR of \citet{Bilicki2021} instead of using the iteratively-selected brightest galaxy in the group as the central one. However, it cannot be determined which of these two central galaxy assignments is a more faithful estimate for the potential minimum based on the reassignment only. Selecting the correct centre of potential is important because the IGL is expected to be centred on it. If one selection method for the central galaxies is better than the other, we expect to see a difference in the radial mass and light profiles of the groups around the group centres from these two methods. The mass density profile around the correct centre is expected to be more peaked and the surface brightness (SB) profile around the correct centre is expected to have more flux in the outskirts where the diffuse light dominate the total light profile of the BGG+IGL compared to the corresponding mass and light distributions around the misidentified centre. In this section, we explore whether such a difference is present in the radial surface mass density profiles and the radial surface brightness profiles in our GAMA group sample. To connect whether the presence of such a difference in the radial profiles is significant enough to identify a preferred central galaxy selection method, we used a carefully produced miscentred group sample from the Hydrangea groups to compare with the corresponding GAMA group profiles. 

\subsection{Selection of miscentred groups in Hydrangea sample}
\label{ss:misc_sample}
In the GAMA group sample, the initial group catalogue considered the brightest galaxy in the group from the iterative method as the halo centre (explained in Sec. \ref{subsec:bcg_kids_gama}). Using SHMR to select the group centres changed the selection criterion from light to associated halo mass of the galaxy in consideration. In the Hydrangea group sample, however, the FOF groups can be centred unambiguously on the true potential minimum of the halo. We considered these as the `true centres' in our analysis. To mimic the central galaxy selection of the GAMA sample, we picked the brightest galaxy in each group as the `updated centres'. Similar to the GAMA groups, this different selection method picked a different central galaxy (compared to the true centres) only for a small subsample of the total group sample (a maximum of 18\%). We considered this subsample of groups with a different updated centre as our `miscentred sample'. The detailed selection methods we used is as follows.

We identified the possible miscentred group candidates using two separate methods. Firstly, we used the projected $r-$band mock images of the groups described in Sec.~\ref{hydrangea}. The central galaxies in the images are at the corresponding group centre of potentials (true centrals) from the Hydrangea group catalogue (see Sec.~\ref{ss:hydrangea_group_sel}). We ran \sextractor \citep{bertin_arnouts1996} on each of the images to select the bright extended sources and picked the brightest sources from each group image to mimic the observational BGG selection. After matching the positions of these brightest galaxies to the true centrals, we found that 18\% of the groups have a brighter galaxy than its true central within a 1Mpc radial distance from its centre of potential.

\begin{figure}
    	\includegraphics[width=\columnwidth]{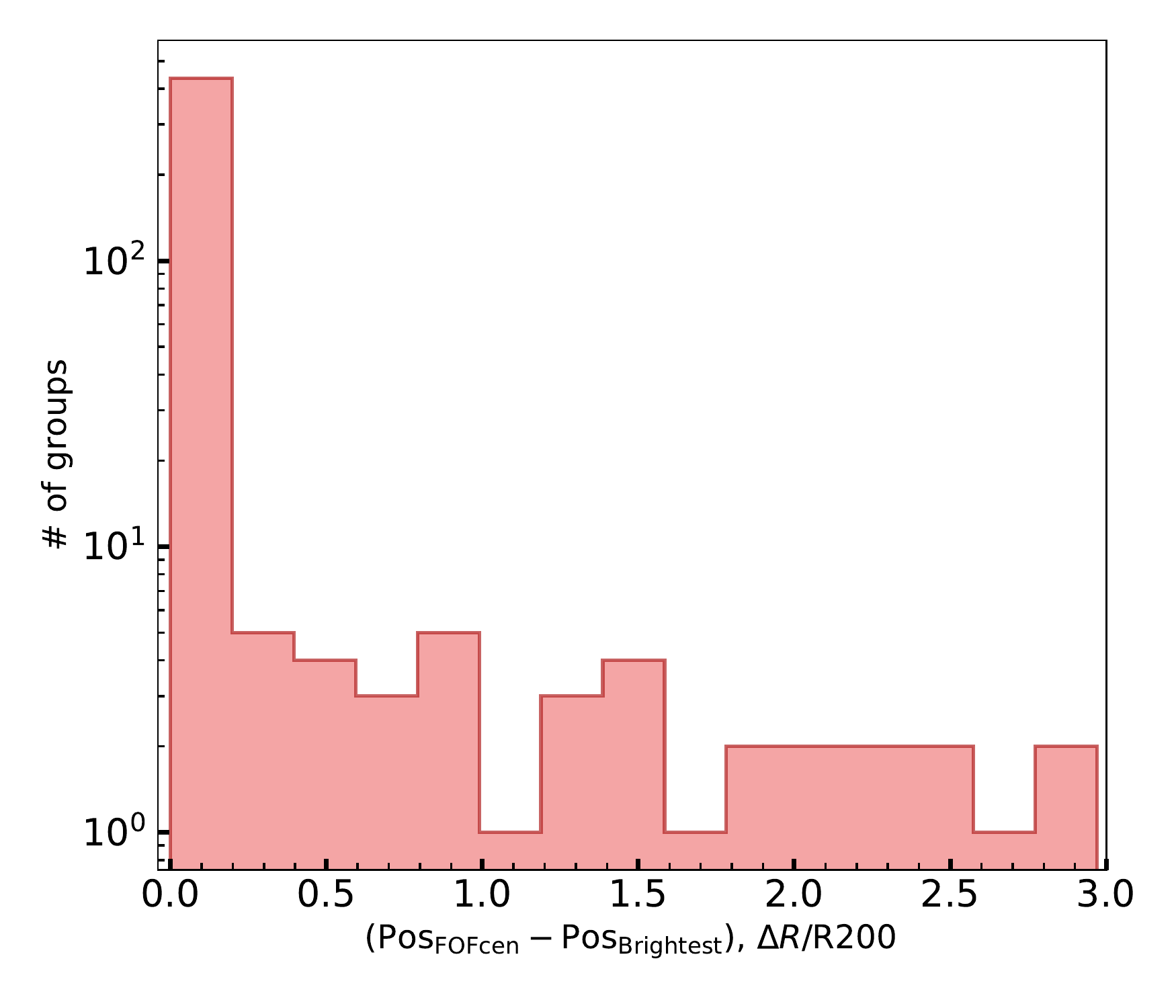}
    	\caption{ {\modtext The distribution of radial distance of the two central galaxy candidates for each of the Hydrangea groups in our sample. The positions of the FOF halo centres were considered as the initial positions and the positions of the brightest group galaxies from the \subfind{} catalogue were considered as the final positions. The distribution is shown in units of the corresponding group $r_{200}$. The y-axis is in logarithmic units.}}
    	\label{fig:hydrangea_delr_miscen}
\end{figure}

Secondly, we directly used the Hydrangea \subfind{} catalogue and picked the brightest galaxy in the $r-$band out of all FOF members as the central galaxy. Here, we considered only the FOF group members as the BGG candidates unlike in the method with the mock images where a projected galaxy can be selected as the brightest galaxy in the group vicinity. After matching the positions with the true centrals, we found that the number of groups having a brighter group galaxy than the true central is reduced to about 10\%. In this second approach, the number of bright galaxies can only go down by eliminating galaxies that only appear to be part of the group in projection, which implies that projections can result in misidentifying the central galaxy. It also suggests that the interlopers can comprise as much as 50\% of the misidentified centrals even in our Hydrangea group sample which do not contain uncorrelated fore-/background galaxies.

There is another important issue related to the FOF group finder that can contribute to the miscentring. Due to the nature of the group finder, it can sometimes merge two smaller groups into one large group if they are not very far away, especially if an in-between galaxy serves as a ``connecting node" \citep[see appendix A of ][]{Jakobs2018}. 
As a result, such groups can exist in the Hydrangea group catalogue and a galaxy that is brighter than the group central can be simply the central galaxy of the second group that got linked to the first group. To exclude such cases, we applied an additional condition that the updated central cannot be more than 500~kpc away from the true group centre. {\modtext A radial distance of 500~kpc is about the same as the average $r_{200}$ of the group sample. The distribution of the distance between the FOF halo centres and the brightest group galaxies are shown in units of $r_{200}$ in Fig. \ref{fig:hydrangea_delr_miscen}. The distribution of physical distances is similar (not shown).} This selection yielded 25 groups (5\%, referred as the clean miscentred sample afterwards) from the Hydrangea sample that still have a brighter galaxy than the true central and therefore, can be misidentified as the group centre in a similar sample of observed galaxy groups. Therefore, even with group membership assigned based on perfect spectroscopic redshifts, the line-of-sight projections can lead to a misidentified central galaxy in the observations. This finding provides a fundamental limitation that needs to be accounted for in a stacking analysis.

\subsection{Mass density profiles of GAMA and Hydrangea groups}
\label{ss:mass_density_profile}
If one selection of the central galaxy is physically preferred to the other, then this is expected to be visible through a more peaked density profile around the better-estimated group centres. To test whether the updated central sample for the GAMA groups is more likely to reside in the group centre of potential, we estimated the stacked stellar mass density profile of the groups around the centrals from the original GAMA catalogue and the BGGs from our updated catalogue. The stellar mass was obtained directly from the StellarMassesLambdarv20 catalogue \citep{Taylor2011,Wright2016}. For the radial distribution of the mass, we calculated the projected distance of each group galaxy to the corresponding group central and normalized it using the corresponding $r_{200}$. We took the total stellar mass of the galaxies in each consecutive radial bin from a stack of all the galaxies in the group sample and divided it by the total surface area of the corresponding annulus to obtain the surface mass density profile. The radial bins did not include the central 5\% of the $r_{200}$ for each group so that the central galaxy was excluded from the density profile. All the 2385 groups in our GAMA group sample were used to calculate the average surface mass density profile. 

\begin{figure*}
    	\includegraphics[width=1.95\columnwidth]{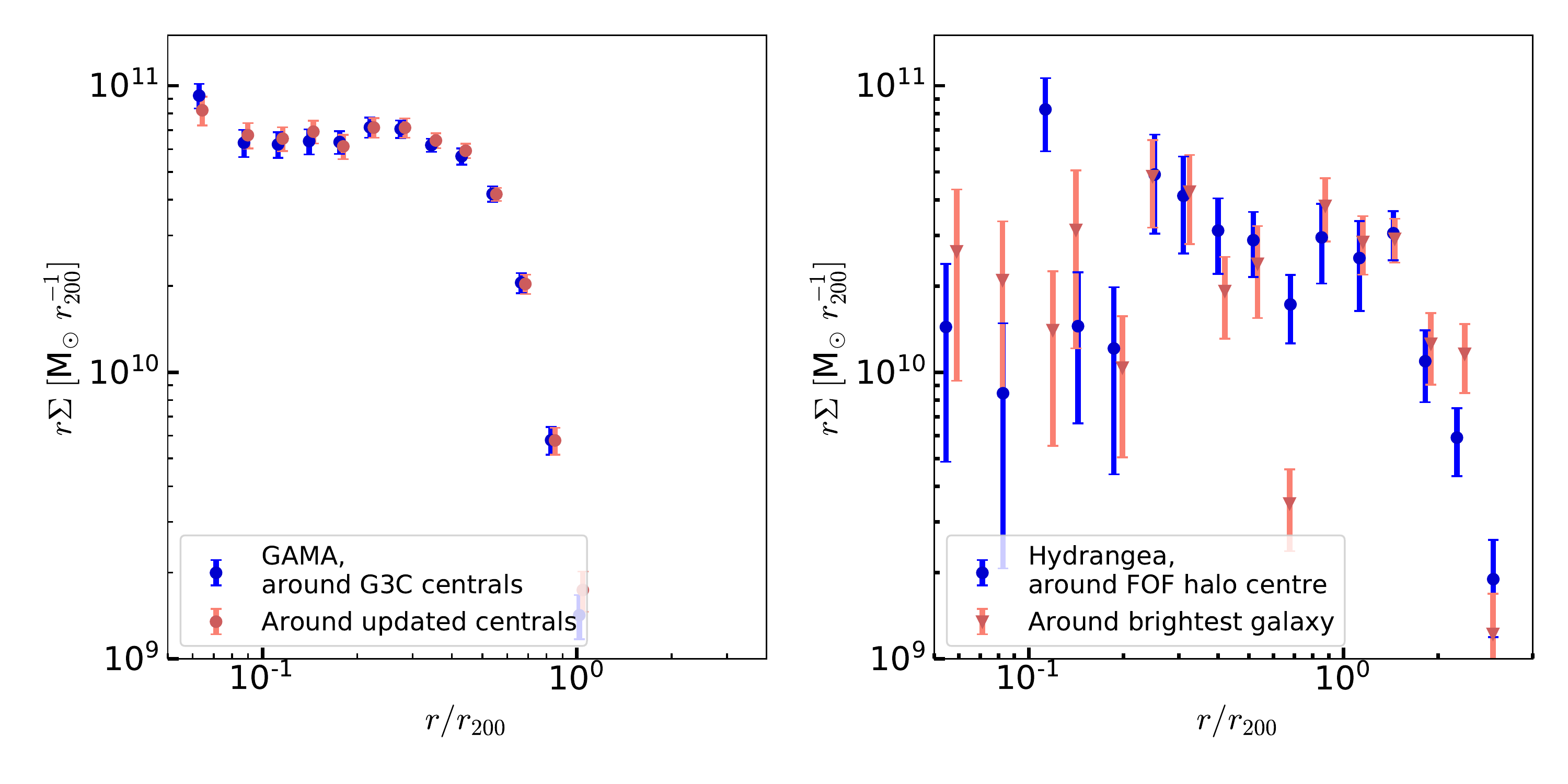}
    	\caption{The total stellar surface density profiles of the stacked GAMA groups where the central candidate was updated (left), and the stacked Hydrangea groups (right, only those with an ambiguous central) around the true central galaxy from the catalogue (blue), and around the brightest group galaxy within 500~kpc distance from the true central galaxy (red). Error bars indicate the 68\% uncertainty from 100 bootstrap re-samplings with replacement. For clarity, the blue and red data points are offset slightly along the x-axis in both panels. For both GAMA and Hydrangea, the profiles are insensitive to the choice of central. {\modtext The re-assignment of central galaxies from G3C centrals to the galaxies with the highest associated halo mass (in GAMA) is not the exact inverse of the re-assignment of central galaxies from FOF halo centres to the brightest group galaxies (in Hydrangea). However, G3C centrals are broadly analogous to the brightest group galaxies from Hydrangea and galaxies with highest halo mass from GAMA are broadly analogous to the FOF halo centres from Hydrangea.}}
    	\label{fig:gama_hydrangea_density_profile}
\end{figure*}

We did not see any significant difference in the stacked profiles of all groups before and after the central galaxy reassignment. To focus on the effect on the groups that are affected by updating the centrals, we also considered only those groups that have been assigned with a new redder central galaxy compared to a bluer one in the original catalogue. We prepared the surface mass density profile for this subsample of 498 groups in the same procedure as described above. The left panel of Fig.~\ref{fig:gama_hydrangea_density_profile} shows the surface density profile of the possibly miscentred groups around the centrals from the GAMA catalogue (blue) and around the centrals from the updated catalogue (red). The error bars on both the blue and red data points show the corresponding 68\% uncertainties and are obtained by 100 bootstrap re-samplings from the respective stack of galaxies with replacement. Even in this case, the mass density profiles around the centrals from the original and updated catalogue do not show any visible difference. The density profiles of the same group samples were also measured using the weak lensing signal of the groups instead of the galaxy stellar mass from the StellarMassesLambdarv20 catalogue, which also did not show any visible difference. 

A possible reason for this lack of difference can be that both of these selections are equally meaningful. In other words, there is a chance that about half of these updated centrals are correctly updated as the central galaxy, whereas the other half were already correctly identified in the original catalogue. 
One reason for this is that, even with spectroscopically selected group members, there is a chance of projection effect from foreground groups. This is not possible to test further with only observational data as there is no information about which galaxies are truly residing at the centre of potential of the groups. Therefore, we use the Hydrangea group sample and their mock $r-$band images to look into this issue in more detail. As shown in Sec. \ref{ss:misc_sample}, we found that about half of the miscentred groups from Hydrangea sample are indeed coming from the LoS projections. Therefore, we repeated the test of the radial mass density profile with the Hydrangea groups before and after taking out the projected subsample from the Hydrangea miscentred sample described in Sec. \ref{ss:misc_sample}.

We prepared the stacked surface mass density profiles around the true centrals and the brightest galaxies (separately) with our clean sample of miscentred groups that were selected as described in Sec. \ref{ss:misc_sample}. The right panel of Fig.~\ref{fig:gama_hydrangea_density_profile} shows the density profile of the Hydrangea group sample (only those with an ambiguous central) around the true central galaxy (blue) and around the brightest galaxy within 500~kpc distance from the true central (red). The error-bars are obtained by 100 bootstrap re-samplings of the galaxies with replacement and show the 68\% uncertainties on the data points. A slight offset was added to the red and blue data points along the x-axis in both the panels for an easier distinction between them. The large uncertainties of the density profile from Hydrangea sample resulted from the significantly smaller miscentred sample size (25) compared to the GAMA miscentred sample (498). The halo mass distribution of the miscentred samples are also slightly different -- {\modtext the GAMA sample has a higher average halo mass which is likely due to the presence of miscentred groups from projection effects.} This resulted in the difference in the normalization of the density profiles on the panels. However, the red and blue profiles on the right panel look almost identical within the error-bars. This behaviour is the same as the GAMA miscentred groups where we were not sure about the true centres of the groups. Therefore, the mass density profiles do not point to a clear physical preference for either of the two centring methods, based on the SHMR or galaxy brightness. 


\subsection{Surface brightness profiles of Hydrangea groups}
\label{ss:SB_profile}
{\modtextt The surface brightness (SB) profile around a misidentified centre is expected to be higher in the central region by definition of selecting the brightest galaxy. On the outskirts, the IGL is expected to be suppressed compared to the IGL around the true centre of potential because the true central galaxy will be treated as a satellite galaxy and masked, and the surroundings of the misidentified centre will only have part of the IGL.} To test whether our group re-centering based on the colour-dependent SHMR improved the central galaxy selection, we looked into the SB profiles of the miscentred Hydrangea group sample. We took the 25 Hydrangea groups from Sec.~\ref{ss:mass_density_profile} that have a brighter galaxy (which is also located within 500kpc of the group centre of potential) compared to their true central. We prepared mock $r$-band images of these groups centred around the updated BGGs (brightest galaxies in the corresponding groups in $r$-band) instead of around their true centrals (centre of potentials) with the same resolution and noise level as we did previously (explained in Sec.~\ref{hydrangea}). These images also span 1Mpc around the updated centrals.

\begin{figure*}
    	\includegraphics[width=2\columnwidth]{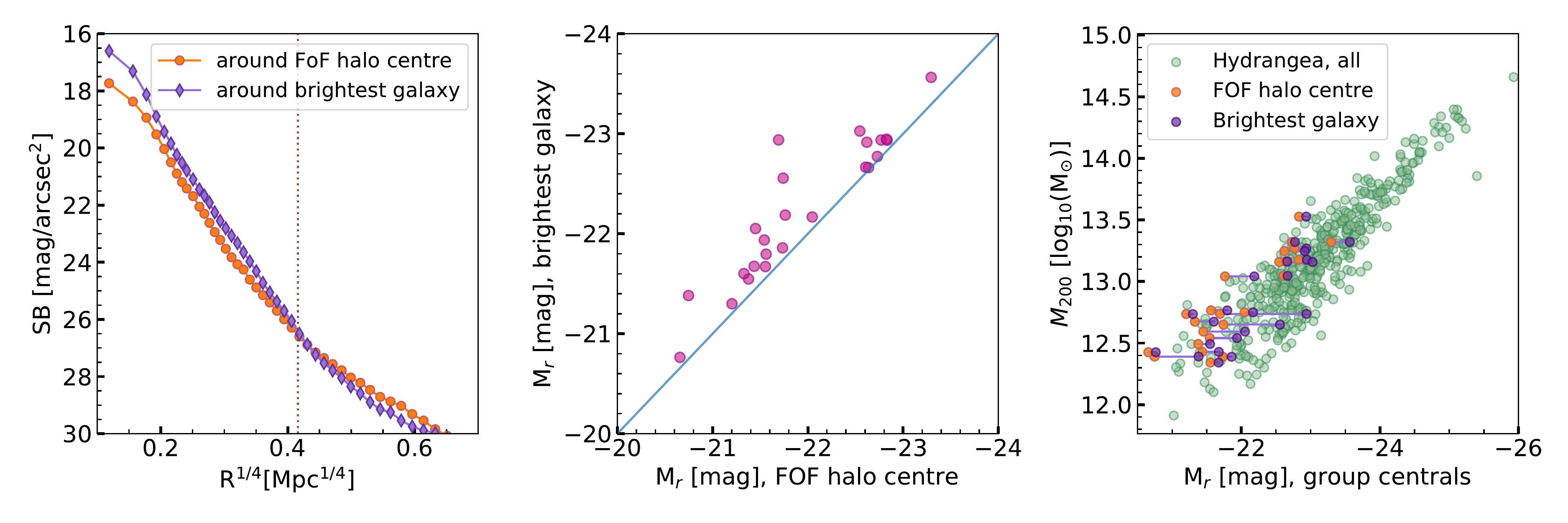}
    	\caption{\emph{Left:} stacked surface brightness profiles of the Hydrangea groups with ambiguous centres around the potential minimum from the catalogue (orange circles), and around the brightest group galaxy within 500~kpc from the true central galaxy (purple diamonds). The profiles centred on the potential minima have lower central brightness compared to those centred on the brightest galaxy. In contrast, the surface brightness of the diffuse IGL (at $R \gtrsim 30$ kpc; vertical dotted line) is higher when the potential minimum is chosen as the centre. \emph{Middle:} absolute $r-$band magnitude of the brightest galaxy (y-axis) vs the absolute $r-$band magnitude of the galaxy at the potential minimum (x-axis) for the subsample of 25 groups in which they are not coincident (pink circles). The blue solid line here shows the 1-to-1 relation. \emph{Right:} halo mass to absolute $r$-band magnitude distribution of the entire Hydrangea sample (light green). The orange and purple points show the miscentred subsample in this parameter space with absolute $r-$band magnitudes of the true and updated BGGs, respectively. {\modtext The FoF halo centre and brightest galaxy from each of the miscentred group are connected with light purple lines.} Groups with lower halo mass and a less luminous BGG are more likely to have a misidentified central as they can have brighter galaxies close to the group centre.
    	}
    	\label{fig:hydrangea_miscen_sb_prof}
 \end{figure*}

We computed two stacked radial SB profiles for all the 497 groups: one centred around the true centrals from the catalogue and the other including the updated BGGs for the 25 miscentred groups. For this, we masked out all the sources except for the central galaxies in the group images so that only the central galaxy and the extended IGL around it remains. Similar to our findings from the density profiles, we found no appreciable difference between the two stacked profiles (not shown here). To focus on miscentring, we then prepared the stacked SB profiles of only the 25 miscentred groups around their true and updated centrals. The left panel in Fig.~\ref{fig:hydrangea_miscen_sb_prof} shows the stacked surface brightness profile around the true centrals in orange and around the updated centrals in purple. 

Two features differentiate the SB profiles around the true FOF centre (orange) and the brightest galaxy as the new group centre (purple). Firstly, the central region of the profile around the updated BGGs (purple) is up to $\approx 1$ mag brighter than the profile around the true centrals (orange). This is also demonstrated by the pink data points in the middle panel of Fig.~\ref{fig:hydrangea_miscen_sb_prof}, where the absolute $r-$band magnitudes of the updated BGGs (y-axis) are brighter by the same amount (the solid blue line shows the position of equal magnitudes for reference) than the true centrals (x-axis) for the 25 miscentred groups. Secondly, even though the profile centred on the brightest galaxy is much brighter at the centre, it is fainter than the profile around the true halo centre beyond 30~kpc (vertical line), where the IGL is dominant. This characteristic agrees with our assumption that around the misidentified central galaxies, the IGL is suppressed. Because the IGL is not evenly distributed around a misidentified BGG, the azimuthally averaged surface brightness around the brightest galaxy is lower than around the actual potential minimum beyond 30~kpc. {\modtext To check whether this suppression is dependant on the luminosity of the central galaxies, we divided the miscentred groups in two bins based on the $r-$band magnitude of the BGGs, and plotted the SB profiles similar to the left panel of Fig. \ref{fig:hydrangea_miscen_sb_prof} (not shown here). We found that the overall behavior that the extended light is suppressed around the brightest galaxy is present in both of the cases, albeit a bit more pronounced in the bin with brighter BGGs.} 

We also checked the distribution of different properties for the groups that are likely to have such misidentified central galaxies. The distribution of the groups across two key parameters, the halo mass ($M_{200}$) and the absolute $r-$band magnitude ($M_r$) of the centrals are shown in the right panel of Fig.~\ref{fig:hydrangea_miscen_sb_prof}. The light green circles show the distribution of all the Hydrangea groups, and the orange and purple circles show the distribution of the true and updated centrals of the 25 miscentred groups, respectively. {\modtext For ease of matching, the FoF halo centre and brightest galaxy from each of the miscentred group are connected with light purple lines.} The points clearly indicate that the less massive groups ($\leq 10^{13.5} \msun$, with a majority $\leq 10^{13} \msun$) in our sample with a less-luminous central galaxy ($\leq-23$ mag, with a majority $\leq-22$ mag) are more likely to have a misidentified central. The points also show that for this work, the majority of our group sample lies above the $10^{13} \msun$ and $-22$ mag threshold. Although the fraction of groups affected by the miscentring is small (about 5\%), to draw any conclusion on the effect of this miscentring on the IGL measurement, we need to quantify the effect. We measure the consequence of this IGL suppression due to the misidentified centrals on the total IGL fraction estimation in Sec.~\ref{ss:bcg_icl_sep}. 

The fraction of miscentred groups (i.e. those for which the brightest galaxy is not the one at the potential minimum) compared to the total number of groups in the halo mass range where we see them in the Hydrangea sample ($\leq 10^{13.5} \msun$) is 7\% (and 15\% if we include the cases where the BGG was a projected galaxy). The fraction is 23\% lower (15\% considering the projected ones) compared to the same fraction from \citet[][their fig. 10]{lange2018} in the same halo mass range. Working with the GAMA group sample only, \citet[][]{olivaAl2014} reported between 10-15\% miscentred centrals at $M_h \leq 10^{13.5} \msun$. An indication to the reason of why the BGGs are not at the halo centre can be found by looking at the magnitude gap ($\Delta m_{12}$) between the BGG and the second brightest galaxies in these groups. \citet{olivaAl2014} found a lower $\Delta m_{12}$ (< 1.0 mag) which suggests a recent halo merger for these groups. In our miscentred sample, all the BGGs except one have $\Delta m_{12}$\ < 1.0 mag, implying a recent halo merger being the most likely reason for the central galaxy to not be the brightest galaxy in the group.

To summarize, we found a noticeable difference in the SB profiles around the true centres and the brightest galaxies in the groups. Therefore, studying the SB profiles can be a way to identify the true central galaxy in groups. {\modtext In this work, we could only test the SB profiles for the simulated sample, which showed IGL suppression at the outskirts of the SB profiles, and therefore demonstrated that a halo-mass based central galaxy selection is more accurate compared to selecting the brightest galaxy as the central one.} We also found that for our sample, the most plausible reason behind the miscentring is interlopers and recent halo mergers. 

\section{Towards a better IGL interpretation}
\label{sec:icl_on_bcg_morphology}
\subsection{Effect of group and central galaxy properties on the stacking}
\label{sec:bgg_prop_on_icl}

To study the effect of group and central galaxy properties on the IGL measurement and to test whether there is a quantitatively preferred way to stack multiple groups that makes the physical interpretation of the IGL more straightforward, we computed the azimuthally averaged radial surface brightness profiles for each of the 497 Hydrangea groups (including the central clusters for a wider halo mass range). For this purpose, we took the mock $r-$band images described in Sec.~\ref{hydrangea}, and ran \sextractor to identify all the bright sources. Starting from the \sextractor segmentation map, we created masks to remove all the sources apart from the central galaxy. To eliminate any residual extended light from the satellites, we extended all the masks by 5 pixels ($\approx 1''$). The mask thus selects only the light from the central galaxy and extended IGL around it in each image. Following a similar method as \citet{zibetti2005}, we fitted the inner region of the surface brightness profiles with a single-component de Vaucouleurs profile \citep[][SD from now on]{devaucouleurs1948} to separate the central galaxy from the extended IGL. We fitted the profile out to $0.2 \times r_{200}$ to completely encompass the central in our group sample with a varied halo mass (and hence $r_{200}$) range. Using the fitted profile as a model for the central, we subtracted it from the masked SB profile to obtain the IGL profile in the outskirts. Also, we obtained the SB profile of the satellite galaxies in the group by subtracting the masked central+IGL profile (azimuthally averaged) from the total group SB profile that was obtained from the unmasked group image. Finally, we integrated these light profiles to obtain the total flux within the central, IGL, and satellites in each group. Thereafter, we calculated the IGL-to-total ($f_{\textrm{IGL}}$), central-to-total ($f_{\textrm{central}}$), and satellite-to-total ($f_{\textrm{satellite}}$) light fractions and examined the distributions of these light fractions with respect to different group properties including the halo mass, central magnitude, richness, and integrated group luminosity. 

\begin{figure*}
	\includegraphics[width=2\columnwidth]{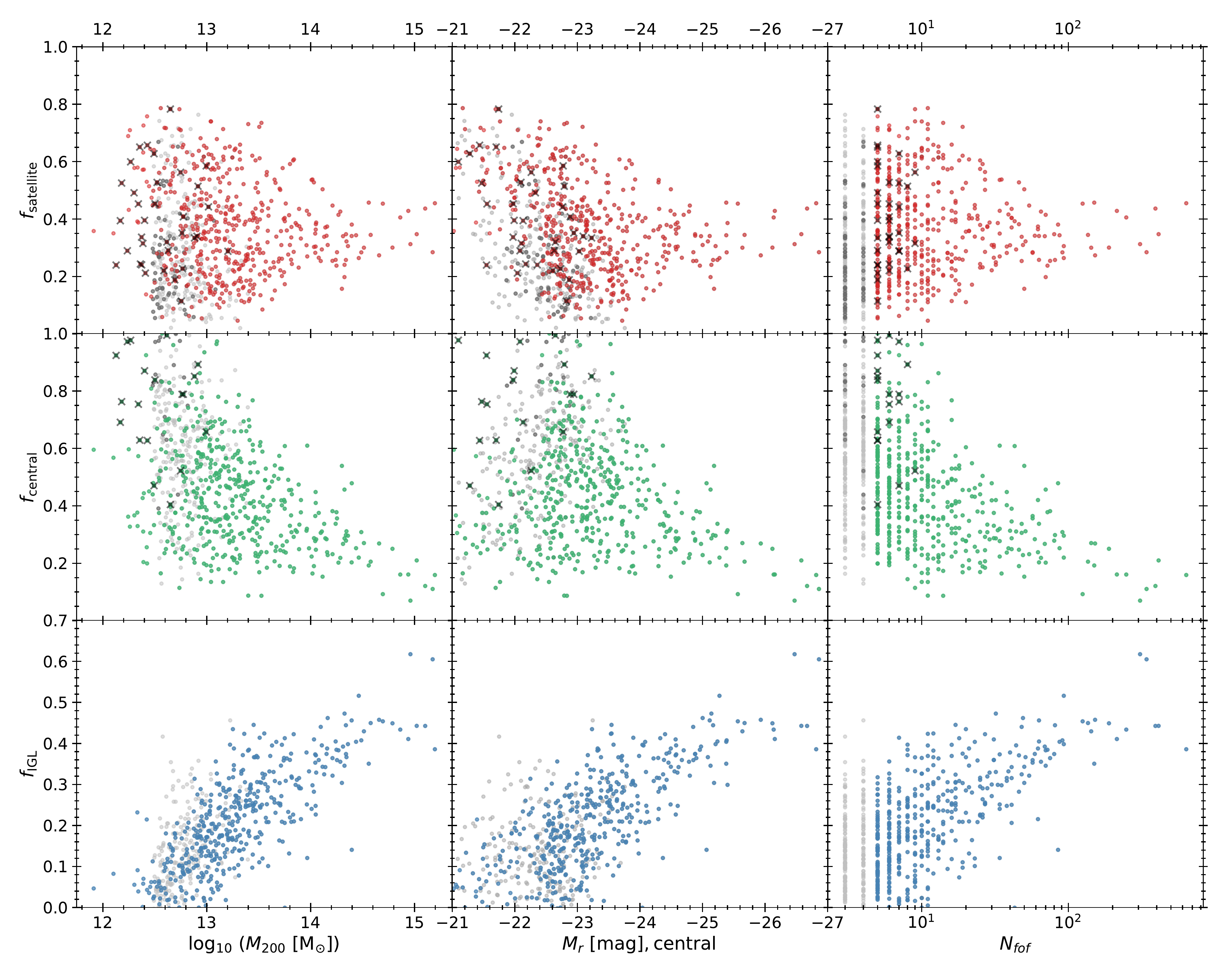}
	\caption{Fraction of the total stellar $r$-band light in satellites (top), the central galaxy (middle), and the IGL (bottom) of Hydrangea groups with respect to, respectively, the group halo mass (left panels), absolute $r$-band magnitude of the central galaxy(middle panels), and group richness (right panels). Black {\modtext crosses} represent parameters resulting from bad fits (see text). Light grey points represent groups with $N_{\textrm{FOF}}<5$ but $M_{200}\geq3\times 10^{12}\msun$. The satellite and central fractions do not show any strong correlations with halo mass, $r-$band magnitude, or group richness (but there is some variation in the scatter). However, the IGL fraction correlates strongly with all of these properties.
	}
	\label{fig:sat_bcg_icl_frac_vs_props}
\end{figure*}

Figure~\ref{fig:sat_bcg_icl_frac_vs_props} shows the distributions of $f_{\textrm{satellite}}$ (top panels), $f_{\textrm{central}}$ (middle panels), and $f_{\textrm{IGL}}$ (bottom panels) for all the Hydrangea groups against the group halo mass ($M_{200}$, left panels), $r-$band magnitude of the group BGGs ($M_\textrm{r}$, middle panels), and group richness ($N_{\textrm{FOF}}$, right panels), respectively. The group richness was measured considering only the galaxies that have an apparent $r-$band magnitude $\leq 19.8$ and stellar mass $\geq 10^{9} \msun$. The red, green, and blue dots represent those groups for which the single-component de Vaucouleurs fitting procedure resulted in a good fit to the inner 45kpc profile and did not exceed the total central+IGL light profile at larger radii. However, $\sim$~6\% of the groups did not satisfy these criteria and led to $f_{\textrm{IGL}}$ having a negative value. In Fig.~\ref{fig:sat_bcg_icl_frac_vs_props}, the black {\modtext crosses} in the top and middle panels indicate these bad fits. In the bottom panel, the bad fits are not visible as they do not have a positive IGL fraction. 

The satellite and central fraction (top and middle row) do not show any strong correlations with $M_{200}$, $M_\textrm{r}$, or $N_{\textrm{FOF}}$. {\modtext The upper envelope of $f_{\textrm{central}}$ distributions appear as monotonically decreasing functions, whereas the lower envelopes show zero correlation. All the panels showing satellite and central fractions have a high scatter in their distribution, especially at the lower mass, lower central luminosity, and lower richness.} The scatter at the highest halo mass, brightest central luminosity, and highest richness ends are smaller and a slight correlation (positive for the satellites and negative for the centrals) is visible there. However, these are the most massive clusters from the Hydrangea sample. The large scatter at the lower mass end indicates that the relative contribution of central and satellite galaxies is highly variable in groups. As groups merge and acquire increasing number of satellites, the satellites may contribute more to the total group light and correspondingly, the central galaxy contributes less to the total group light. Therefore, we conclude that in the group-mass range, the scatter is too high to consider any trend in the top and middle panels.

However, the IGL fraction ($f_{\textrm{IGL}}$) shows a clear trend with all of $M_{200}$, $M_\textrm{r}$, and $N_{\textrm{FOF}}$, as shown in the bottom panel of Fig.~\ref{fig:sat_bcg_icl_frac_vs_props}. A similar trend is also visible in other recent cosmological simulation-based IGL/ICL analyses. \citet{pillepich2018illustris} found a similar $f_{\textrm{IGL}}$ (mass) vs $M_{200}$ relation at $10^{13}\leq m_{200}/\msun \leq 2\times 10^{14}$ with a radial selection method to separate the central and ICL in the IllustrisTNG simulations. Using a similar ICL separation method as \citet{pillepich2018illustris}, \citet{ContiniGu2021} also found a positive ICL mass vs host halo mass correlation. By using a phase-space-based galaxy finder algorithm to separate the host galaxies from what they refer to as the `intra-halo stellar component' (IHSC), \citet{Canas2020} reported a positive correlation of the fraction of IHSC with the host $M_{200}$ and $N_{\textrm{FOF}}$ for haloes with $10^{11}\leq m_{200}/\msun \leq 10^{13}$ as well. However, \citet{Montes2022} found no significant correlation between the $f_{\textrm{IGL}}$ with the host $M_{200}$ after combining multiple observational studies of the ICL measurements at $z<0.07$ (see their fig. 3), which also had a large scatter in the $f_{\textrm{IGL}}$ between 5-40\%. They speculated that the lack of correlation between the ICL fraction and the halo mass may indicate that both groups and clusters have similarly efficient ICL formation mechanisms. With a contrary argument, a large scatter in the observational data can be the result of the different systematics in the individual data sets from the different studies. 

Before discussing the dependence of IGL fractions in our sample in further detail, it is worth mentioning that we have tested a range of further properties and found no significant correlation with any of the satellite, central, or IGL fractions. These properties include the average rest-frame $u-r$ colour of the centrals, the effective radii (the radius encompassing 50\% of the total light of the component) of the central and IGL components, and the integrated $r-$band magnitude of the group. 

In all the panels, the grey points show groups with $N_{\textrm{FOF}}<5$, but within a comparable mass range as the main sample ($M_{200}\geq3\times10^{12} \msun$). In the left and middle columns, these points follow the trend of the coloured points, implying that a halo-mass based group selection gives similar results compared to a richness-based one.

Returning to the bottom panels of Fig.~\ref{fig:sat_bcg_icl_frac_vs_props}, the similar correlation of $f_{\textrm{IGL}}$ with $M_{200}$, $M_\textrm{r}$, and $N_{\textrm{FOF}}$ is expected as these are not mutually independent properties. More massive and richer groups have a higher chance of accumulating more mass and light in the central galaxy and the surrounding region. We see, however, a larger scatter in $f_{\textrm{IGL}}$ with respect to $N_{\textrm{FOF}}$, especially for $N_{\textrm{FOF}} \leq 10$ in the bottom right panel of Fig.~\ref{fig:sat_bcg_icl_frac_vs_props}, compared to the tighter relations with $M_{200}$ and $M_\textrm{r}$. A possible reason for the scatter with $N_{\textrm{FOF}}$ being larger can be the presence of fossil groups, which are defined as the relics of old galaxy groups where the central galaxy grows predominantly by merging with satellite galaxies that are at least as luminous as the characteristic luminosity of the galaxy luminosity function for the system \citep{ponman1994}. In fossil groups, $N_{\textrm{FOF}}$ is low even though the mass and luminosity of the central galaxy are high enough to be comparable to a rich group or even a cluster with a correspondingly high IGL fraction. An opposite scenario may also occur in fossil groups where the central galaxy has gone through a recent merger, and hence the IGL is not in place yet, which will result in a smaller IGL fraction. 

A detailed analysis of whether the groups in our sample are potentially fossil groups is beyond the scope of this paper. However, a quick test of looking into the magnitude gap between the BGG and the second brightest galaxy in the groups with $N_{\textrm{FOF}} \leq 10$ showed that about 18\% of these groups have at least a magnitude gap of 2 in $r-$band, and for 8\% of these groups, the second brightest galaxy is also located within $0.5\ r_{200}$ distance from the BGG. According to the widely used definition of fossil groups by \citet{jones2003}, a system must have a minimum X-ray luminosity of $10^{42}\ h^{-2}_{50}\ \textrm{erg s}^{-1}$ and a $r-$band magnitude gap $\geq2$ between the two brightest group members that are within $0.5\ r_{200}$ distance from each other. The X-ray luminosity limit corresponds to a minimum halo mass of about $10^{13}\msun$ \citep{stanek2006} which, along with the distance and magnitude criteria, selects about 4\% of the low-richness groups in our sample only. Therefore, according to these criteria, about 4\% of the groups in our sample with $N_{\textrm{FOF}} \leq 10$ are likely fossil groups confirming that fossil groups are a minor but non-negligible contributor to the scatter in the IGL fraction at the low-richness end. 

\begin{figure}
	\includegraphics[width=\columnwidth]{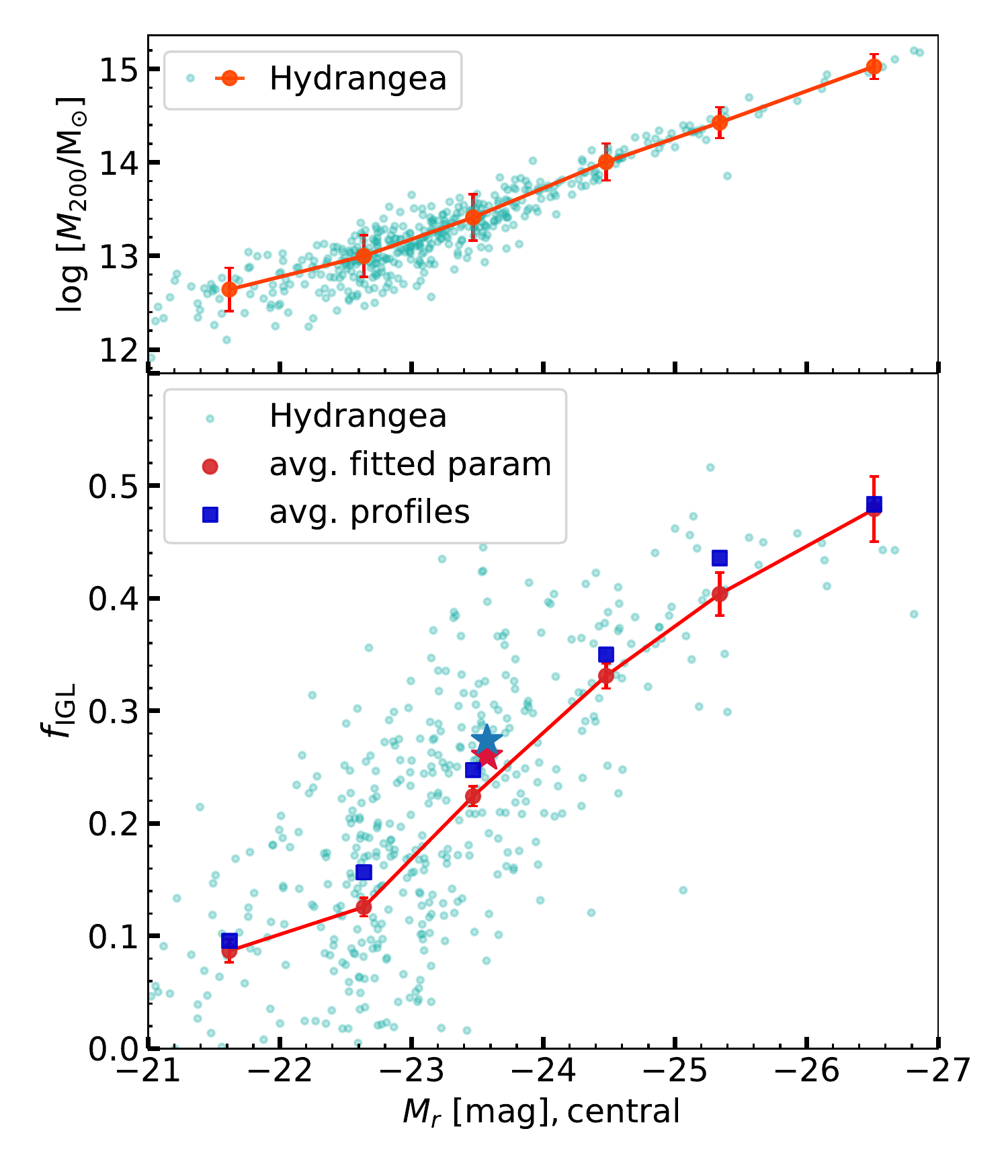}
	\caption{Top: halo masses of the Hydrangea group and cluster sample with respect to their absolute $r-$band central galaxy magnitudes (turquoise). The orange circles and error bars indicate the mean values and standard deviations of the distribution in each magnitude bin. Middle: fraction of IGL to the total group light with respect to the absolute $r-$band central galaxy magnitudes of the Hydrangea group sample. The turquoise points indicate the IGL fraction for each individual group obtained by subtracting the central galaxy with a single de Vaucouleurs (SD) fit to the central 40 kpc. Red circles indicate the average values (magnitude-weighted) of individual fits in 1~mag wide bins in central galaxy $M_\textrm{r}$. Error bars indicate $1\sigma$ uncertainties on the mean. The deep blue squares show the IGL fraction obtained by stacking the individual surface brightness profiles of the group in the same magnitude bins, and then fitting them using an SD profile to subtract the central galaxy. The close agreement between the two methods indicates that stacking the surface brightness profiles of the groups gives a nearly unbiased estimate of the IGL fraction. The red and blue stars indicate the entire ensemble average of the individual $f_{\textrm{IGL}}$, and the $f_{\textrm{IGL}}$ from the stack of all the galaxy profiles considered in this plot, respectively.}
	\label{fig:icl_frac_vs_lr}
\end{figure}

Unlike the halo mass, $M_\textrm{r}$ of the central galaxy is directly observable. Therefore, its correlation with $f_{\textrm{IGL}}$ is more straightforward to test from the observational data. Because of this, we look into the strong trend of $f_{\textrm{IGL}}$ with $M_\textrm{r}$ in Fig.~\ref{fig:icl_frac_vs_lr}. The middle panel here is based on the $f_{\textrm{IGL}}$ vs $M_\textrm{r}$ plot from Fig.~\ref{fig:sat_bcg_icl_frac_vs_props} and shows $f_{\textrm{IGL}}$ of the Hydrangea groups as the turquoise data points. To identify the average trend of $f_{\textrm{IGL}}$ with respect to $M_\textrm{r}$, we binned the groups along $M_\textrm{r}$ from $-21\textrm{ mag}$ to $-27 \textrm{ mag}$ and calculated the mean $f_{\textrm{IGL}}$ in each bin. The mean values are obtained from the halo mass-weighted average of the individual data points to account for the fact that more massive groups will have a higher fraction of IGL contribution in the stacked profile. The red circles and solid line in the middle panel show the average $f_{\textrm{IGL}}$ values in each magnitude bin. Each bin is plotted at the average $M_\textrm{r}$ of its centrals. The error bars associated with the red points show the statistical $1\sigma$ uncertainty on the mean. 

To test whether this trend is also present in the SB profiles and is not only an outcome of any bias in the fitting procedure of the individual group profiles, we stacked the surface brightness profiles of the Hydrangea groups in the same $M_\textrm{r}$ bins as the red data points. After fitting the average surface brightness profiles of the $M_\textrm{r}$ bins and obtaining $f_{\textrm{IGL}}$ following the same procedure as before, we obtained the data points shown by the deep blue squares in the middle panel of Fig.~\ref{fig:icl_frac_vs_lr}. The similarity between the stacked and individually measured IGL fractions has two important implications. Firstly, the trend of increasing $f_{\textrm{IGL}}$ with respect to the $M_\textrm{r}$ of the group centrals is confirmed. The top panel of Fig.~\ref{fig:icl_frac_vs_lr} shows the correlation between the group halo mass with the $M_\textrm{r}$ of the group centrals in turquoise data points. The orange circles show the average halo mass in the equivalent $M_\textrm{r}$ bins compared to the middle panel of the figure. The tight correlation of the increasing halo mass with respect to the $M_\textrm{r, central}$ indicates why we see an increased $f_{\textrm{IGL}}$ in the middle panel. More massive groups have a brighter central galaxy which also had a chance to accumulate a larger amount of IGL during its growth. Secondly, this similar trend for both the red and deep blue points indicates that the stacking of the surface brightness profiles preserves the underlying IGL fraction distribution in individual groups. Therefore, stacking the profiles to increase the signal-to-noise ratio in observational data is indeed a valuable tool to measure the faint IGL signal. 

However, {\modtext groups with central galaxy in brighter $M_\textrm{r}$ bins having larger $f_{\textrm{IGL}}$  means that if all the groups are stacked together to obtain the IGL fraction, this trend is not evident. Ignoring this positive trend of IGL fraction with central galaxy luminosity} can bias or limit conclusions that can be drawn from the analysis. The $f_{\textrm{IGL}}$ from the mean SB profile is also likely to be biased towards a slightly higher value because of the higher light contribution from brighter centrals. This is highlighted by the red and blue stars in the middle panel, they indicate the entire ensemble average of the individual $f_{\textrm{IGL}}$, and the $f_{\textrm{IGL}}$ from the stack of all the galaxy profiles considered in this plot against the average galaxy $M_\textrm{r}$, respectively. The red and blue stars are located closely, showing again that stacking the profiles (blue star) preserves the behavior of stacking the fitted $f_{\textrm{IGL}}$ (red star). They also show that the values of such broad average {\modtext (25.6\%$\pm$0.7\%)} are slightly higher than the mean trend of the magnitude-based bins {\modtext (22.4\%$\pm$0.9\%)} at the same magnitude, {\modtext the values in parentheses here are the magnitude weighted average and the standard error to the weighted average of the corresponding samples.} Although this is not a large difference, a $M_\textrm{r}$-based sub-stacking results in a more accurate estimation of the $f_{\textrm{IGL}}$ while preserving the properties of the underlying central galaxy population.
Therefore, instead of stacking a sample of groups with a varied range of central galaxy luminosity, stacking groups in narrow bins of $M_\textrm{r}$ will result in a more straightforward interpretation of the measured IGL fraction. 


\subsection{Central-IGL separation}
\label{ss:bcg_icl_sep}

In Sec.~\ref{sec:bgg_prop_on_icl}, we followed \citet{zibetti2005} and used a single-component de Vaucouleurs (SD) profile fit to separate the central galaxy from the extended IGL from the satellite-masked group image. This method is also used in other studies of IGL, such as \citet{Kluge2021}. {\modtext Another approach to separate the central and IGL is to fit a double de Vaucouleurs profile to the combined central+IGL light profile, one to fit the inner central galaxy, and another to fit the IGL at the outskirts \citep[e.g.][DD method from now on]{gonzalez2005,Kluge2021}. In this method, the SB profile of the central+IGL light was divided into two regions. The regions were simultaneously fitted with separate de Vaucouleurs profiles such that the sum of the fitted profiles has the minimum $\chi^2$ value compared to the group SB profile. The fitting parameter ranges ensured that the two de Vaucouleurs profiles were in the appropriate radial zones for the central galaxy ($\leq 40$~kpc) and IGL ($\geq 40$~kpc). 

While fitting the surface brightness profiles of individual groups, the SD method could successfully separate the central and IGL with a reasonably well-fitted central galaxy profile for $\sim 94\%$ of the cases (discussed in more detail in Sec.~\ref{sec:bgg_prop_on_icl}). The DD method, however, was unsuccessful in more than 80\% of the groups for individual group profile fits. Upon visual inspection, the primary reason for the failure was the presence of additional features in the outer regions of the SB profiles of individual groups.} These irregularities in the light profile are likely caused by either tidal features of the stars in these regions or by local enhancements in the light profile from recent star formation activity. Such local light enhancement from star formation can be more pronounced in the simulations due to the stellar mass resolution of the simulations, which acts as a lower limit to the added light in such regions. The stellar mass resolution ($\sim 10^6 \msun$) in the Hydrangea simulations means that in any region with ongoing star formation, the minimum added stellar mass of young (and bright) stars is of the order of $\sim 10^6 \msun$. This feature gets averaged out near the group centre which is already bright. However, in the outskirts, where the distribution of light is sparse, even one young star particle can cause a significant deviation from a regular de Vaucouleurs profile \citep[see also][]{Trayford2017}. 

These irregularities in the group light profiles are, however, smoothed out during the stacking procedure. Even in the case of sub-stacking based on luminosity bins as discussed in Sec.~\ref{sec:bgg_prop_on_icl} where the number of stacked groups in each bin was not very high ($\leq 100$), the profiles became smoother, and both the SD and DD fitting procedures worked for all the sub-stacked groups.
\begin{figure*}
	\includegraphics[width=\textwidth]{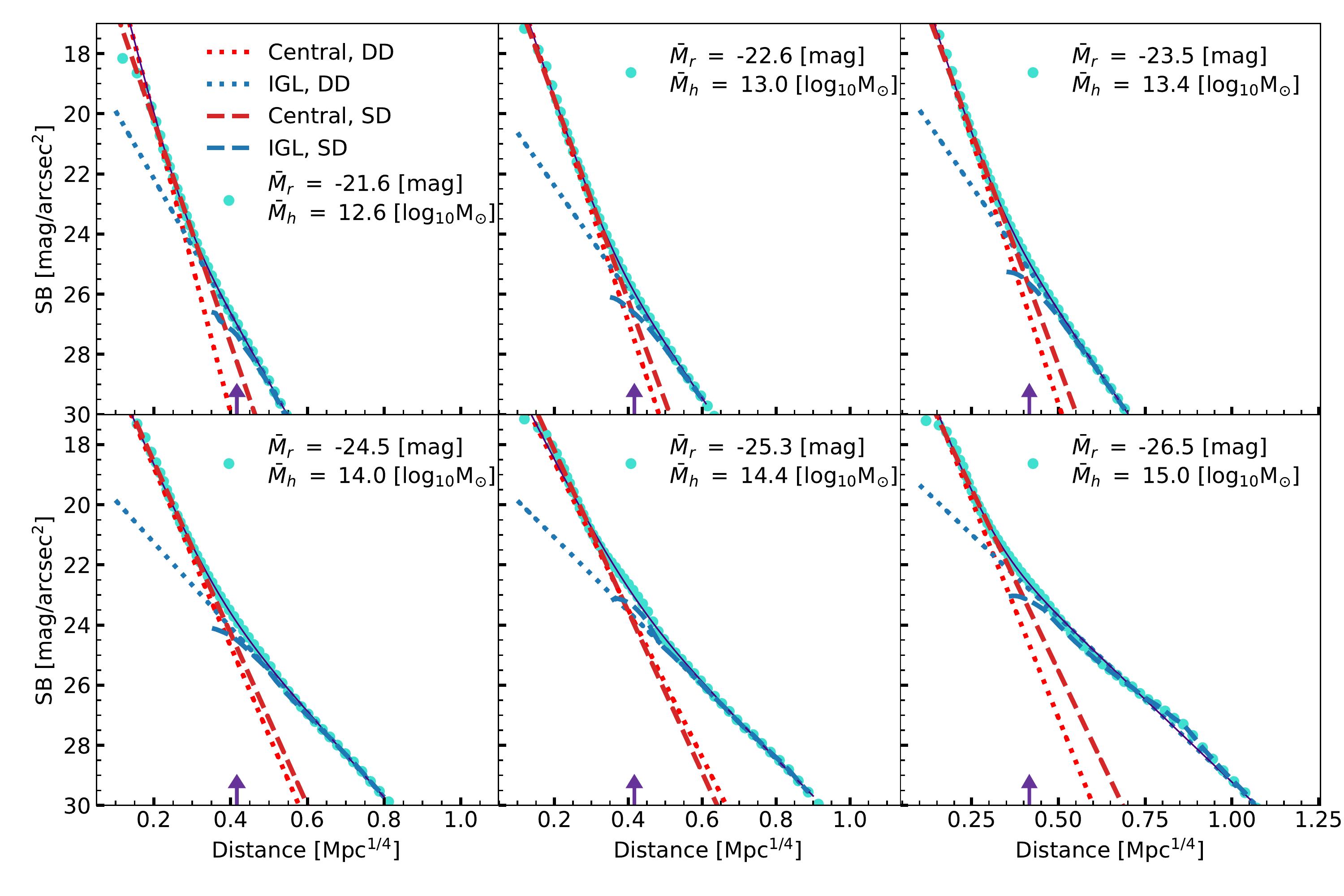}
	\caption{The radial surface brightness profiles of Hydrangea groups stacked within narrow magnitude bins. Turquoise circles show the stacked surface brightness profiles of the Hydrangea groups with the average central galaxy magnitude and average halo mass of the groups indicated in the top-right corner of each panel. Dashed lines indicate the best-fit single de Vaucouleurs profile (SD) of the central (red) and the remaining light as the IGL (blue). The dotted lines indicate the best-fit double de Vaucouleurs (DD) profile of the central (red) and the IGL (blue). The solid purple line shows the total central+IGL light from the double de Vaucouleurs fit. In all bins, the single and double de Vaucouleurs profiles provide comparably good fits to the data.}
	\label{fig:stack_profiles_multiple}
\end{figure*}
Figure~\ref{fig:stack_profiles_multiple} shows the sub-stacked surface brightness profiles of Hydrangea groups based on their corresponding BGG luminosities. In each of the six panels in Fig.~\ref{fig:stack_profiles_multiple}, cyan circles show the surface brightness profiles of the sub-stacked central+IGL. The average values of the absolute $r-$band magnitudes ($\Bar{M}_{\textrm{r}}$) of the centrals and the group halo masses ($\Bar{M}_{\textrm{h}}$) of the groups in the stacks are given in the top-right corner of each panel. The dashed and dotted lines show the profiles from the SD and DD fitting methods, respectively. Red and blue lines show the central galaxy and IGL profiles, respectively, in all panels. The solid purple lines show the combined fitted central+IGL profiles from the DD method. The most prominent feature in this plot is that all the profiles are well-fitted by the SD and DD methods. Also, while the SD and DD fitted lines do not overlap entirely, the red lines indicate similar regions for the centrals, and the IGLs begin to have a higher contribution to the total light compared to the centrals  at a similar radial distance in each subplot. However, due to the way the IGL is defined in these methods, the SD and DD fitted IGL lines cover different radial ranges in all the subplots. We compare the IGL measurement from these SD and DD methods in more detail in Fig.~\ref{fig:icl_frac_sd_dd}.

\begin{figure*}
    \includegraphics[width=2\columnwidth]{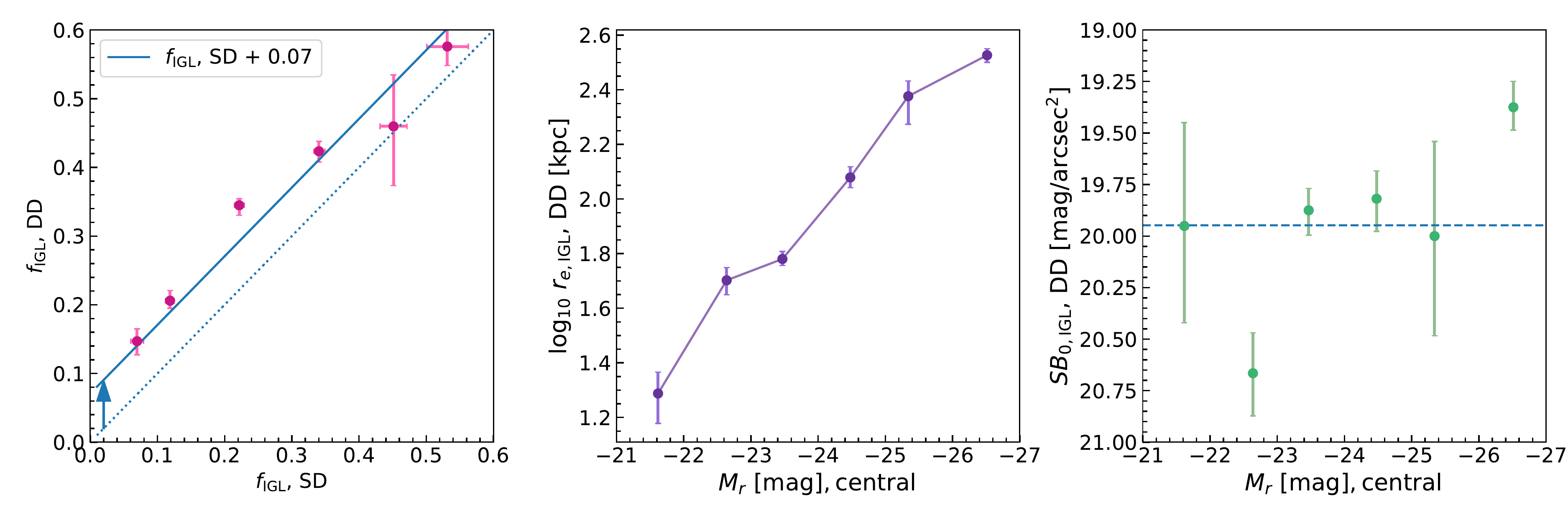}
    \caption{Left: comparison of the IGL fractions obtained from fitting the stacked surface brightness profiles with a single de Vaucouleurs profile (SD, $x$-axis) and a double de Vaucouleurs profile (DD, $y$-axis). The IGL fraction is systematically higher in the DD fitting. The solid blue line is the best linear fit with slope of unity to the individual points, with an offset of +0.07 (dotted line showing the 1-to-1 relation). Middle: effective radius ($r_\textrm{e}$) of the IGL profile from the DD fitting (dotted blue lines from Fig.~\ref{fig:stack_profiles_multiple}) with respect to the average $M_\textrm{r}$ bins of the central galaxy. The monotonically higher values of the $r_\textrm{e}$ towards the right indicate that the IGL profile becomes more extended with increasing central galaxy luminosities. Right: central surface brightness of the IGL profiles from the DD fitting with respect to the average $M_\textrm{r, central}$ bins. There is no systematic trend, albeit with some scatter. In all the panels, the error bars show the 68\% confidence interval for the data points.}
    \label{fig:icl_frac_sd_dd}
\end{figure*}

The left panel of Fig.~\ref{fig:icl_frac_sd_dd} directly compares the IGL fractions obtained from fitting the stacked group surface brightness profiles with the SD and DD methods. In all the three panels in this figure, the error bars show the 68\% confidence interval, and they are obtained by 100 bootstrap resampling of the groups in the stack before fitting the profiles. As visible from the left panel, the IGL fractions are consistently and systematically higher when estimated with the DD fit compared to the SD method. The solid blue line here shows the fitted line with unity slope through the points and it is offset by a value of 0.07. This indicates that using a DD fit instead of an SD fit to measure the IGL fraction results in about 7\% higher IGL fraction in any of the magnitude-based sub-stacks in a similar group sample. By comparing to the fitted IGL profiles from Fig.~\ref{fig:stack_profiles_multiple} (dashed and dotted blue lines for the SD and DD fits, respectively), this excess likely comes from the IGL fraction from the inner region in the DD fitting. Because of the methodology we used to measure the central galaxy, IGL, and satellite fractions in these groups (see Sec.~\ref{sec:bgg_prop_on_icl}), this difference resulted in the central galaxy having a lower light fraction in the DD fits compared to the SD fits. The satellite light fraction, however, remained unchanged by definition. The $\sim$ 7\% difference in the estimated $f_{\textrm{IGL}}$ between the SD and DD methods is similar to what \citet{Kluge2021} reported by using the same two methods to measure the $f_{\textrm{IGL}}$ in their sample of 170 galaxy clusters at $z\leq0.08$. However, their measured $f_{\textrm{IGL}}$ range (13-18\%) is significantly smaller (albeit with a large scatter in the measurement) than what we find (35-50\% on the cluster scale). Using stacking analysis on SDSS groups and SD fitting to separate the central galaxy and IGL, \citet{zibetti2005} also found a smaller $f_{\textrm{IGL}}$ (10.9$\pm$5.0). Based on simulated data and DD fitting, \citet{Puchwein2010} reported a similar $f_{\textrm{IGL}}$ compared to our measurements here. For a more detailed comparison of $f_{\textrm{IGL}}$ measurements among observational and simulation-based studies using different methods, interested readers are referred to table 1 of \citet{Kluge2021}. Comparing these existing works to the left panel of Fig.~\ref{fig:icl_frac_sd_dd} demonstrates that although $f_{\textrm{IGL}}$ can vary depending on the data and the method to separate the IGL from the central galaxy, the scatter of $f_{\textrm{IGL}}$ measurements between different methods can be quantified. These systematic differences among other common methods of ICL measurements can be explored and used for a robust comparison between $f_{\textrm{IGL}}$ measurements from different studies. 

The middle panel of Fig.~\ref{fig:icl_frac_sd_dd} shows the effective radius ($r_e$) of the IGL profile from the DD fitting (dotted blue lines from Fig.~\ref{fig:stack_profiles_multiple}) against the average $M_\textrm{r, central}$ bins. The $r_e$ values have a positive correlation with the increasing central galaxy luminosities, indicating a more extended profile in groups with a brighter central. The right panel shows the central surface brightness of the IGL profiles from the DD fitting with respect to the average $M_\textrm{r, central}$ bins. They show a small range of the central SB of the IGL profiles regardless of the central galaxy luminosities. Combining with the findings from the middle panel, this indicates that the higher IGL fraction in brighter groups is primarily due to more extended diffuse light rather than a self-similar increase in SB across all radii compared to groups with a fainter central galaxy.

Coming back to the effect of group miscentring from Sec.~\ref{ss:SB_profile}, we have also explored the effect of the miscentring on the IGL fraction. We took the stacked SB profiles around the true centre of potentials and around the brightest galaxies for the 25 groups where these differ (shown in the left panel of Fig.~\ref{fig:hydrangea_miscen_sb_prof}). We measured the central, IGL, and satellite light fractions compared to the total group light for both of these stacks following the same procedure as we have used for the analysis in this section. We see a $\sim 20\%$ difference in the central and satellite fractions, with the central fraction increasing and the satellite fraction decreasing in the miscentred sample when measured around the brightest galaxies rather than the actual centre-of-potential. The change in central and satellite fraction is expected as we are swapping the central galaxies in the images with a brighter galaxy. As this is not at the potential minimum of the halo, there are also fewer satellite galaxies within a 1Mpc radius around it in the mock images which can affect the SB profile at larger radii. However, we see almost no change in the IGL fraction (from 14\% to 13\%). This lack of change is likely because the miscentring occurs only for the less massive groups  where the IGL fraction is already small (for majority of such groups, it is <20\%). For such small IGL fractions, the difference of the IGL profiles at the outskirts of the blue and red profiles from the left panel of Fig.~\ref{fig:hydrangea_miscen_sb_prof} can only measure up to around 1\% compared to the total light, which is a much smaller value than the scatter of the IGL fractions from individual groups ($\sim 10\%$). Therefore, the slight difference in the SB profiles does not have any significant impact on the total IGL measurement. This was also reflected by the initial test of the surface brightness profile of all the groups (Sec.~\ref{ss:SB_profile}) where we did not notice any visible difference in the profiles around the true and updated centrals. Moreover, almost all the ambiguous centrals were found to contain signatures of a recent halo merger, which means that these bright galaxies are most likely central galaxies of another group halo that is falling into the host halo in consideration. In that case, these alternative centrals may well have their own IGL around them that they assembled before the merger. Therefore, we can expect that $\sim 5\%$ possibly misidentified centrals will not add any significant bias in the detection and analysis of the IGL in our observed GAMA group sample (to be presented in a companion study, Ahad et al., in prep.).

\subsection{Radial (u-r) colour profile of central+IGL}
\label{ss:bcg_icl_color_age_metallicity}
Different formation mechanisms of the IGL are expected to leave a distinct imprint in the stellar populations of the IGL, which can be traced by IGL properties such as colour and metallicity. For example, the age, colour, and metallicity profiles of the IGL are very different if the dominant formation channel of IGL is tidal stripping of massive satellites, total disruption of dwarf galaxies, or stellar ejection after major mergers. The gradient of the profiles can indicate which mechanisms were dominant, and the intrinsic values of the properties can point to which type of galaxies contributed to the IGL the most \citep{Montes2022,Contini2021}. Multiple studies based on groups and clusters have found negative gradients in the radial colour profile of the central+IGL which can indicate a gradient in either age or metallicity \citep[more details can be found in][and references therein]{Montes2022,Contini2021}. However, the age-metallicity degeneracy makes disentangling these two quantities from colours alone a challenging endeavour. Having data from a bluer photometric band such as the $u-$band can potentially help with resolving this issue. As we have information about the real age, metallicity, and colour from the simulations, here we studied the radial distribution of these three properties for our group sample.

\begin{center}
    \begin{figure}
    	\includegraphics[width=\columnwidth]{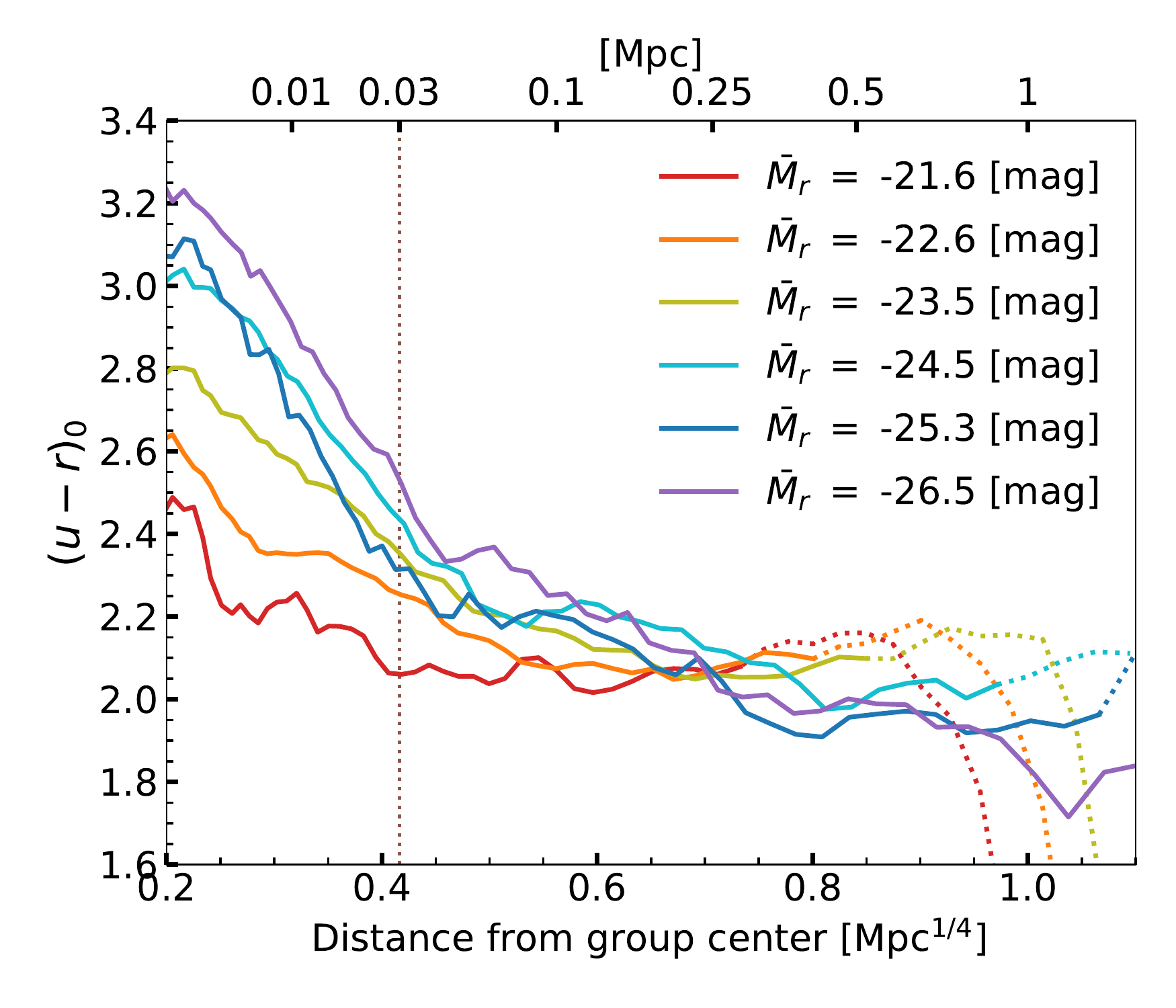}
    	\caption{The radial profile of the rest frame $u-r$ colour of the central+IGL from Hydrangea groups in different central magnitude bins. Each line represents a different central galaxy magnitude bin as shown in the top right corner. The brown dotted vertical line at 30~kpc from the group centre indicates the typical extent of the central galaxy. All the lines are solid until the average $r_{200}$ within their bin, and dotted beyond.}
    	\label{fig:ur_rad}
    \end{figure}
\end{center}

Figure~\ref{fig:ur_rad} shows the radial $(u-r)_0$ colour profiles of the central+IGL in our central magnitude-based sub-stacks of the groups. We obtained the azimuthally averaged $(u-r)_0$ colour profiles from the mock images of the groups in $u$ and $r$ bands which include only the star particles that are part of the central group galaxy and IGL. The different colours in the figure indicate different magnitude bins (mean magnitudes of the bins are shown in the top right corners). 
Out to 30~kpc, the approximate extent of the central galaxy (indicated by the vertical dotted line), all the colour profiles have a negative gradient which is stronger for the brighter centrals. At larger radii, the difference in gradient between the low and high mass groups (with fainter and brighter centrals, respectively) becomes more prominent. The profile of groups with fainter centrals (red and orange) have a shallower or flatter profile compared to the ones with a brighter central (blue and purple) which show a stronger negative gradient out to at least 500~kpc. It may indicate that compared to the lower mass groups, the light in central+IGL of the higher mass groups grew preferably by accretion over mergers. But to confirm any such indication, it is necessary to look into the age and metallicity profiles of the same sample.  

\begin{center}
    \begin{figure*}
    	\includegraphics[width=2\columnwidth]{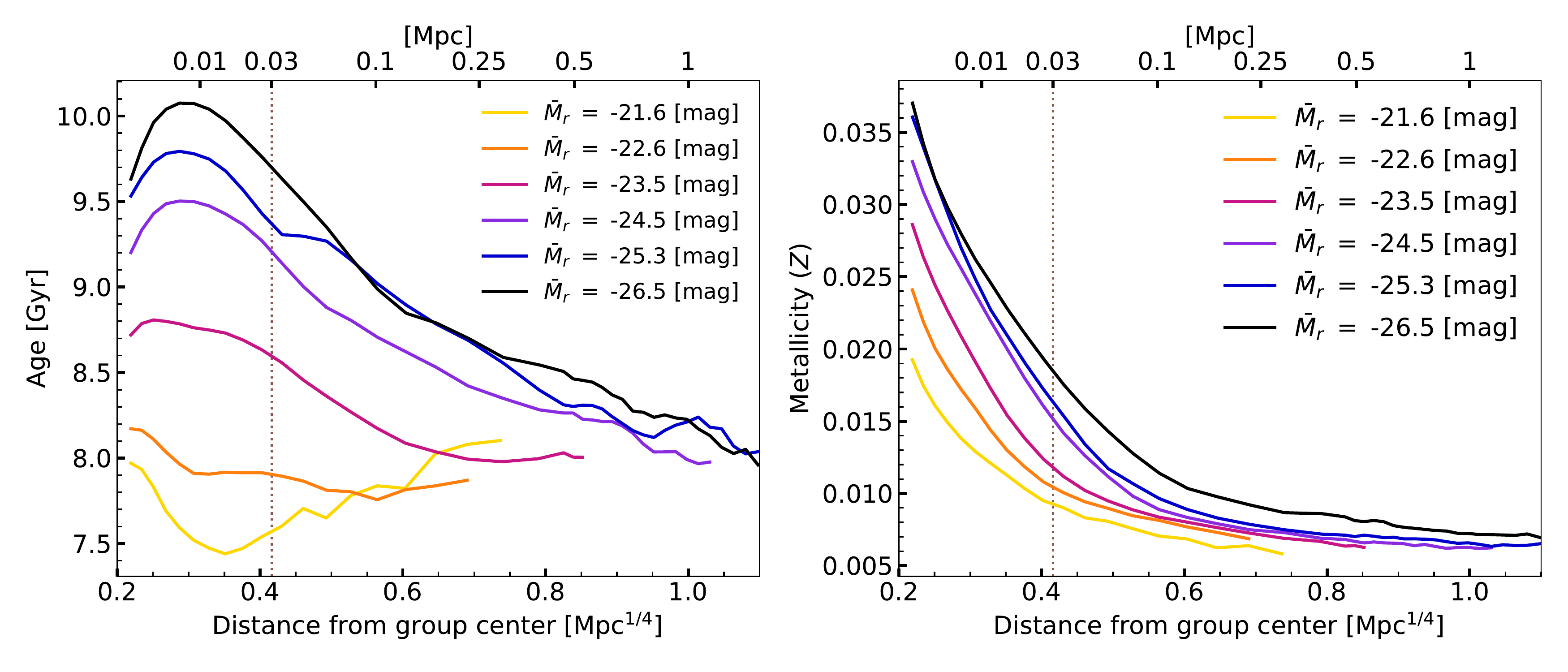}
    	\caption{The radial age (left) and metallicity (right) profiles of the central+IGL from Hydrangea groups in different central galaxy magnitude bins. In both panels, the brown vertical dotted line shows 30~kpc distance from the group centres. The age profiles show large variation between the groups with faint and bright centrals. The metallicity profiles have similar trends for all the magnitude bins. }
    	\label{fig:age_met}
    \end{figure*}
\end{center}

We explore the age and metallicity profiles of our sample in Fig. \ref{fig:age_met}, computed directly from the simulation outputs. The left and right panels show the azimuthally averaged radial profiles of the age and metallicity of the group central+IGL in the same magnitude bins as Fig. \ref{fig:ur_rad}, respectively. The different colours indicate the different magnitude bins and the mean magnitudes of the bins are shown in the legends. For both the panels, the brown vertical dotted line shows a 30~kpc distance from the group centres. For higher mass groups and clusters (with brighter centrals), the radial age profiles follow a similar negative gradient pattern to the $u-r$ profiles from Fig. \ref{fig:ur_rad}. The lower mass groups (fainter centrals, yellow, orange, and magenta) have a flatter or irregular age profile which is comparable to the corresponding $u-r$ colour profiles of the corresponding groups. For the metallicity profiles, all the bins have a similar negative gradient in the inner part (<100~kpc) that approaches a flatter profile in the outskirts. 

In all three profiles (colour, age, metallicity), the visibly different intrinsic values for different magnitude bins within 30~kpc from the centre indicate that brighter centrals are on average redder, older, and have a higher metallicity which is expected in general for giant elliptical galaxies \citep[e.g.][]{peletier1990,davis1993,Huang2018,Santucci2020}. However, the colour profiles here trace the metallicity profile more strongly than the age. The profiles beyond 30~kpc here indicate a different story for different central magnitude bins (or different halo mass ranges). The high mass groups/clusters (M$_r<-24$) have a similar negative age gradient from the centre to the far outskirts. In this case, the colour profiles are more correlated to the age profiles compared to the metallicity profiles. The shallower metallicity profiles in the outskirts are an indication that the IGL was built via the accretion of satellite galaxies. The stellar population of the accreted galaxies then got mixed to produce the flat metallicity distribution \citep[e.g.][]{Montes2021}. The negative gradient of the age profile at the outskirts of the high mass groups, together with the similar colour profile, also supports the scenario of IGL growth via accretion. In contrast, the flatter colour profile at the outskirts of the lower mass groups may indicate ICL formation through expelled stars from a major merger \citep{krick2007,Demaio2018}. The flat age profile of the low mass groups with a younger average age also supports this hypothesis. However, it is worth mentioning that the intrinsic scatter of these average age profiles for the lower mass groups is rather high ($\sim$ 1~Gyr, not shown here). This is likely resulting from a combination of multiple IGL formation scenario as different groups can have different IGL formation histories which is averaged out in a stacked sample like ours. A smaller intrinsic scatter for the higher mass groups ($\sim$ 0.2 Gyr), on the other hand, indicates that the IGL formation history is more homogeneous for such groups. However, the stacks in narrow magnitude bins still retain some key signatures of the corresponding group ensemble. 

{\modtext To sum up, the negative gradient of the age and colour profiles at the outskirts of the high mass groups indicate the scenario of IGL growth via accretion, whereas the flatter age and colour profiles of the low mass groups support the scenario of IGL formation via major mergers.} Therefore, we conclude that in our sample, lower mass groups likely accumulated their IGL predominantly from mergers and the higher mass groups/clusters likely accumulated their IGL from the accretion of stripped stars from the outskirts of other galaxies.

\section{Summary and Conclusions}
\label{sec:conclusions}
The growth and extent of the diffuse light in galaxy groups (IGL) can provide important insights to improve our understanding of hierarchical cosmic structure formation. However, identifying the centre of a galaxy group in an unambiguous way is observationally nontrivial and may introduce non-negligible bias in the IGL measurements. Central galaxy candidates in galaxy groups have a galaxy colour bimodality in their stellar-to-halo-mass relation (SHMR) that can lead to a misidentification of group centres from only stellar mass (or luminosity) based BGG selection. In this paper, we investigated the effect of misidentifying the group centre on the IGL measurements using data from the Hydrangea cosmological hydrodynamic simulation suite and a comparable group sample from the GAMA survey. We also studied the dependence of the IGL on the properties of the central galaxies in the groups. Our main findings are summarised as follows:

\begin{itemize}
    \item Using a galaxy colour-based SHMR \citep{Bilicki2021} to identify the central group galaxies instead of the luminosity-based selection of the \citet{Robotham2011} group catalogue leads to a re-assignment of the group centre in 23 per cent of the GAMA groups in our sample. Applying a similar procedure to mock images of the Hydrangea groups, we found that nearly the same fraction (18 per cent) of them do not have their r-band brightest galaxy at the potential centre of their halo, out of which 49 per cent are due to line of sight projection. The rest are most likely resulting from recent halo mergers.
    \item Despite the difference in the selected central galaxy, the radial galaxy-mass density profiles remain unchanged when centring on the true halo centre (for Hydrangea) or the updated central galaxy (for GAMA) instead of the brightest group galaxy (BGG). However, the surface brightness (SB) profile of the Hydrangea groups is suppressed by up to 0.5 mag beyond 30~kpc (where IGL dominates) when centred on the BGG rather than the true group centre. Therefore, a similar difference in the SB profile can indicate miscentring in central group galaxy selection. But such miscentring does not significantly affect the {\modtext measurements of IGL fraction}, because even amongst ambiguously centred groups, this suppression only affects the total IGL fraction measurement by $\sim$1 per cent.  
    \item To separate the central galaxy from the IGL, we fitted a single-component de Vaucouleurs (SD) profile to the inner 40~kpc of the satellite-excised SB profile. The excess above the fitted central galaxy light was identified as the IGL. The estimated IGL fraction out of the total group light is positively correlated with the host halo mass, central galaxy magnitude, and richness, albeit with a higher scatter at the low richness end. This correlation indicates that during a stacking analysis to measure the IGL, a sub-stacking based of the central galaxy magnitude will make the interpretation of the measurements more straightforward.
    \item We also used a double de Vaucouleurs (DD) fit to separate the central and IGL in the SB profile. However, the DD fitting was unsuccessful for 80\% of the individual groups, which was primarily due to the presence of additional features in the SB profile beyond 50~kpc from the centres. Such features are smoothed when we stacked the groups in narrow magnitude bins, and both the SD and DD fitting worked well for the stacked profiles. The estimated IGL fractions using an SD fitting compared to a DD fitting to the central+IGL light are consistently about 7\% lower for all the magnitude bins. This difference can be used to calibrate IGL measurements from different studies that use either of these methods. Such systematic differences among other common methods can be explored for a robust comparison among different IGL measurement studies.
    \item The central surface brightness of the IGL from the stacked DD fits is very similar for all the magnitude bins, and the half-light radius of the IGL from the DD fit gets larger from fainter to brighter centrals. This finding indicates that groups with brighter centrals have more extended IGL.
    \item The rest-frame $u-r$ colour, age, and metallicity profiles of the central+IGL are different for different magnitude bins, with brighter centrals being redder, older, and more metal rich at a given radius. This suggests that the dominant IGL formation channel for the low mass groups is likely major mergers, whereas the dominant IGL formation channel for the high mass groups/clusters is likely stellar stripping from satellite galaxies.
\end{itemize}

The findings from this work will be used to interpret the IGL component in the GAMA galaxy groups using KiDS imaging data and can be used as crucial predictions to interpret the IGL component from the upcoming next-generation survey data like \textit{Euclid} or LSST.

\section*{Acknowledgements}

{\modtext We thank the reviewer for valuable comments that helped to improve the presentation of our results.} We thank Ivan K. Baldry for providing constructive comments on the manuscript. 

SLA and HH acknowledge support from the Netherlands Organization for Scientific Research (NWO) under Vici grant number 639.043.512. YMB acknowledges support from NWO under Veni grant number 639.041.751. The authors thank Shun-Sheng Li for sharing measurements of the weak lensing signal around the GAMA groups with an ambiguous centre. 

The Hydrangea simulations were in part performed on the German federal maximum performance computer ``HazelHen'' at the maximum performance computing centre Stuttgart (HLRS), under project GCS-HYDA / ID 44067 financed through the large-scale project ``Hydrangea'' of the Gauss Center for Supercomputing. Further simulations were performed at the Max Planck Computing and Data Facility in Garching, Germany. This work used the DiRAC@Durham facility managed by the Institute for Computational Cosmology on behalf of the STFC DiRAC HPC Facility (www.dirac.ac.uk). The equipment was funded by BEIS capital funding via STFC capital grants ST/K00042X/1, ST/P002293/1, ST/R002371/1 and ST/S002502/1, Durham University and STFC operations grant ST/R000832/1. DiRAC is part of the National e-Infrastructure.

This research made use of data from the Galaxy and Mass Assembly survey (GAMA). GAMA is a joint European-Australasian project based around a spectroscopic campaign using the Anglo-Australian Telescope. The GAMA input catalog is based on data taken from the Sloan Digital Sky Survey and the UKIRT Infrared Deep Sky Survey. Complementary imaging of the GAMA regions is being obtained by a number of independent survey programs including GALEX MIS, VST KiDS, VISTA VIKING, WISE, Herschel-ATLAS, GMRT, and ASKAP providing UV to radio coverage. GAMA is funded by the STFC (UK), the ARC (Australia), the AAO, and the participating institutions. The GAMA website is \href{http://www.gama-survey.org}{http://www.gama-survey.org}.

The analysis of this work was done using Python (\href{http://www.python.org}{http://www.python.org}), and Jupyter notebook \citep{Kluyver2016jupyter}, including the Python packages \textsc{NumPy} \citep{Harris_et_al_2020}, \textsc{AstroPy} \citep{astropy2013}, and \textsc{SciPy}
\citep{jones2009}. Plots have been produced with \textsc{Matplotlib} \citep{hunter2007matplotlib}. 

\section*{Data Availability}

The data presented in the figures are available upon request from the corresponding author. The raw simulation data can be requested from the C-EAGLE team \citep{bahe2017hydrangea,barnes2017cluster}.



\bibliographystyle{mnras}
\bibliography{Diffuse_light_in_galaxy_groups} 

\bsp	
\label{lastpage}
\end{document}